\documentclass[aip, amsmath,amssymb,reprint,nofootinbib]{revtex4-1}

\usepackage{color}
\usepackage{graphicx}% Include figure files
\usepackage{graphicx,epstopdf}
\usepackage{dcolumn}% Align table columns on decimal point
\usepackage{bm}% bold math
\usepackage{mathrsfs}
\usepackage{soul}
\usepackage{mathtools}
\usepackage{blkarray, bigstrut}
\usepackage{hyperref}

\usepackage{mathtools}

\usepackage{graphics}
\usepackage{amssymb}
\usepackage{amsmath}
\usepackage{mathtools}
\usepackage{color}

\usepackage{graphicx}
\usepackage{epsfig}
\usepackage{multirow}
\usepackage{epstopdf}
\usepackage{subfigure}
\usepackage{mathtools}
\usepackage{hyperref}
\usepackage{graphicx}
\usepackage[space]{grffile}
\usepackage{latexsym}
\usepackage{textcomp}
\usepackage{longtable}
\usepackage{tabulary}
\usepackage{booktabs,array,multirow}
\usepackage{amsfonts,amsmath,amssymb}
\usepackage{epstopdf}
\usepackage{comment}
\usepackage[table]{xcolor}
\usepackage{notoccite}
\usepackage{graphicx}% Include figure files
\usepackage{dcolumn}% Align table columns on decimal point
\usepackage{bm}% bold math
\usepackage{graphicx}% Include figure files
\usepackage{xcolor}
\graphicspath{{./figures/}}
\usepackage{hyperref}% add hypertext capabilities
\usepackage{amsfonts}
\usepackage{amssymb}
\usepackage{amsmath}
\usepackage{notoccite}
\usepackage{lineno}
\usepackage[utf8x]{inputenc}
\usepackage{lineno}
\usepackage[T1]{fontenc}
\usepackage{mathptmx}
\usepackage[inline]{enumitem}

\begin{document}

\title{Controlling species densities in structurally perturbed intransitive cycles with higher-order interactions} 

%\author{Sourin Chatterjee$^{1,*}$, Sayantan Nag Chowdhury$^{2,*,\dagger}$, Dibakar Ghosh $^{2}$, and Chittaranjan Hens$^{2,\dagger}$}
%
%\address{$^{1}$  Indian Institute of Science Education and Research,
%	Kolkata, West Bengal 741246, India \\
%	$^{2}$ Physics \& Applied Mathematics Unit, Indian Statistical Institute, B T Road, Kolkata 700108, India\\
%	$^{*}$ Both the authors have contributed equally to the work\\
%	$^{\dagger}$ Corresponding author.
%	E-mail addresses: chittaranjanhens@gmail.com (Chittaranjan Hens), jcjeetchowdhury1@gmail.com (Sayantan Nag Chowdhury)}	

%\title{Emergent rhythms in coupled nonlinear oscillators due to dynamic interactions}

\author{Sourin Chatterjee$^{*}$}

\affiliation{Department of Mathematics and Statistics, Indian Institute of Science Education and Research,
Kolkata, West Bengal 741246, India}

\author{Sayantan Nag Chowdhury$^{*}$}
\email{jcjeetchowdhury1@gmail.com}

\affiliation{Physics and Applied Mathematics Unit, Indian Statistical Institute, 203 B. T. Road, Kolkata-700108, India}

\affiliation{Technology Innovation Hub (TIH), IDEAS (Institute of Data Engineering Analytics and Science Foundation), @ Indian Statistical Institute, 203 B. T. Road, Kolkata-700108, India}

\author{Dibakar Ghosh}

\affiliation{Physics and Applied Mathematics Unit, Indian Statistical Institute, 203 B. T. Road, Kolkata-700108, India}

\author{Chittaranjan Hens}
\email{chittaranjanhens@gmail.com}

\affiliation{Physics and Applied Mathematics Unit, Indian Statistical Institute, 203 B. T. Road, Kolkata-700108, India}

\affiliation{International Institute of Information Technology, Gachibowli, Hyderabad-500032, India}

\date{\today}

\begin{abstract}\footnote{$^{*}$ Both the authors have contributed equally to the work.}
	
	The persistence of biodiversity of species is a challenging proposition in ecological communities in the face of Darwinian selection. The present article investigates beyond the pairwise competitive interactions and provides a novel perspective for understanding the influence of higher-order interactions on the evolution of social phenotypes. Our simple model yields a prosperous outlook to demonstrate the impact of perturbations on intransitive competitive higher-order interactions. Using a mathematical technique, we show how alone the perturbed interaction network can quickly determine the coexistence equilibrium of competing species instead of solving a large system of ordinary differential equations. It is possible to split the system into multiple feasible cluster states depending on the number of perturbations. Our analysis also reveals the ratio between the unperturbed and perturbed species is inversely
proportional to the amount of employed perturbation. Our results suggest that nonlinear dynamical systems and interaction topologies can be interplayed to comprehend species' coexistence under adverse conditions. Particularly our findings signify that less competition between two species increases their abundance and outperforms others.

\end{abstract}

\maketitle

	\begin{quotation}
	{\bf  Recent trends in ecological research suggest the pairwise interaction among competitors may not readily infer the mechanism of tremendous biodiversity of species. Due to this difficulty, ecologists often resort to the structure of diverse competitive higher-order networks. Recently, Grilli et al.\ \cite{grilli2017higher} shed light on the stability of biodiversity in higher-order networks, while pairwise competition allows only the neutral cycling around the stationary points. We study the influence of structural perturbations on the intransitive cycle of higher-order species interactions and attempt to capture the unexplained complexities of competitive ecosystems. Our analytical treatment provides a straightforward method to determine the globally stable stationary point without solving a large number of ordinary differential equations with higher-order terms. A tiny structural perturbation among the competitors in the higher-order structure leads to the cluster synchrony, where the unperturbed model settles into the complete synchronized death state. Our study suggests that two species can easily enhance their respective densities in comparison to others by reducing the pairwise winning probability from the stronger species. Our simple mathematical framework with the presence of intransitive competition and higher-order interactions advances our understanding of biodiversity maintenance among competitive communities.}
	
\end{quotation}

\section{Introduction}

Understanding how species interact, compete, and try to push each other to extinction while preserving bio-diversity is a long-standing problem in ecology. As a result, several theoretical and statistical approaches to replicating biodiversity have been developed \cite{lotka1920analytical,volterra1926variations,may1975nonlinear,chowdhury2021complex,hubbell2011unified,may1974biological,chowdhury2021eco,may2019stability,chesson2000mechanisms,nag2020cooperation,gao2016universal,hens2019spatiotemporal,roy2022multigames}.

%According to neutral theory, species are rare or
%abundant not because of their traits and the traits of their
%competitors but rather solely because of stochastic drift in
%densities of competitively identical species (11).
%However, both of the classical theories have limitations and achievements \cite{wennekes2012neutral,chisholm2010niche,gravel2006reconciling,mcgill2003test}. %For instance,  assumptions in neutral theory does not meet with  the reality \cite{mcgill2003test}. 
%For instance, 'neutral drift' becomes inconsistent for the stability of forest diversity \cite{clark2003stability}. Also in competition trade off models,  the paradox in plankton world where number of coexisting species  exceeds the number of limiting resource \cite{huisman1999biodiversity} cannot be explained.
One of the classical approaches is {\it Niche theory}, which predicts that in the same niche, when compared to a strong competitor, the weak one is removed due to the competitive exclusion principle, but large species-coexistence is maintained due to habitat heterogeneity and appropriate adaptation and exploitation of niche for each species \cite{chase2009ecological,soberon2009niches,tilman2004niche}.   
% On the other hand, in {\it neutral theory}, for a given trophic level all the species are ecologically equivalent, i.e all have equal probability to colonize in open space \cite{bell2000distribution,brokaw2000niche} %Brian-o-nell-youtube} 
%and  bio-diversity emerges as stochastic balance between the speciation and extinction \cite{hubbell2011unified}. Neutral theory asserts that there will be overlapping niches.
In contrast, in {\it neutral theory}, all species are ecologically equivalent for a given trophic level, i.e., all have an equal probability of colonizing in open space \cite{bell2000distribution,brokaw2000niche}, and bio-diversity emerges as a stochastic balance between speciation and extinction \cite{hubbell2011unified}. According to neutral theory, there will be overlapping niches. 
However, both classical theories have limitations as well as accomplishments \cite{wennekes2012neutral,chisholm2010niche,gravel2006reconciling,mcgill2003test}. For example, 'neutral drift' is incompatible with the stability of forest diversity \cite{clark2003stability}. Also, the paradox in the plankton world, where the number of coexisting species exceeds the number of limiting resources \cite{huisman1999biodiversity}, cannot be explained by competition trade-off models.

\begin{figure*}[!t]
	\centerline{\includegraphics[width=0.85\textwidth]{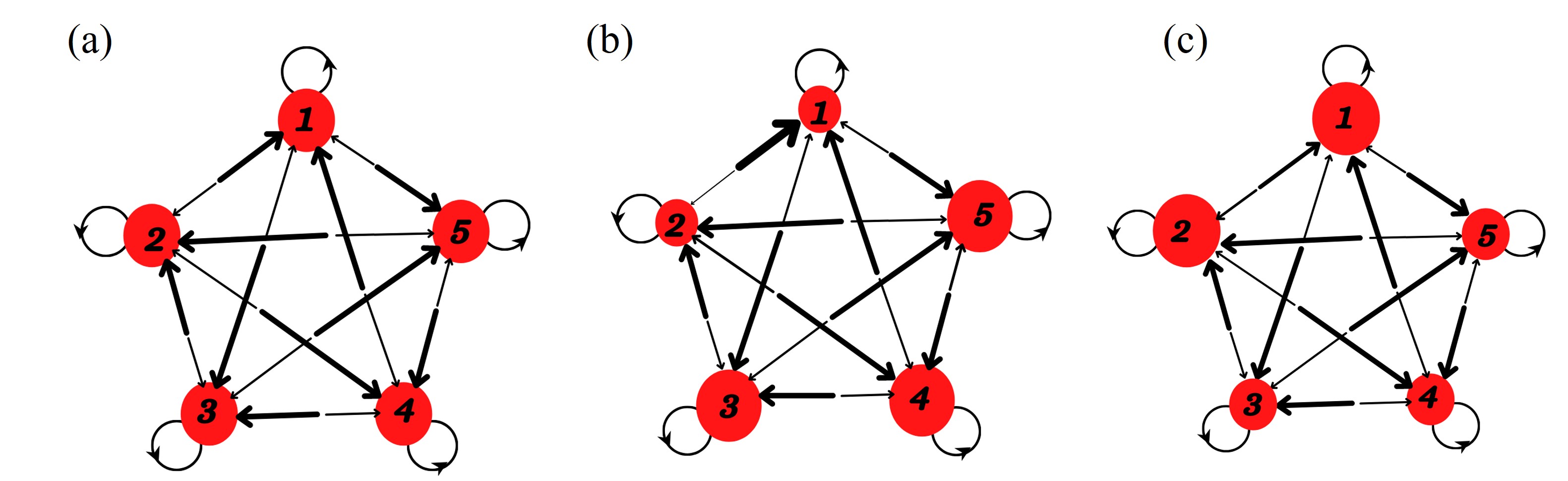}}
	\caption{{\bf Schematic representation of the impact of perturbation}: The red circle denotes species, and its size indicates the density of that species. The two-sided arrow represents the interaction between two species, and its width contemplates the winning probability of the species towards which it is pointed. The self-loop depicts the intra-species interaction with a $50\%$ winning probability. In (a), every species has the same density. In (b), we increase the winning likelihood of species 1 (Stronger species in the interaction) against species 2, the density of both species decreases (size of the circles are decreased compared to species 3,4, and 5). In (c), we increase the winning probability of species 2 (Weaker species in the interaction) against species 1, the density of both species increases (size of the circles are now enhanced compared to others). }
	\label{prac11}
\end{figure*}

\par
Motivated by the shortcomings of classical theories, people are looking for alternative frameworks, particularly those based on game theory. One of them is the widely studied intransitive cyclic competition \cite{he2011coexistence, szabo2007evolutionary,berr2009zero,sinervo1996rock,kerr2002local,laird2006competitive, hibbing2010bacterial,gilpin1975limit}, in which bio-diversity is maintained within a community despite strong competition. Even in the presence of hierarchy, this type of formalism allows a large number of species to converge, particularly into stable cycles or, in some cases, into a stable equilibrium. This phenomenon is known as the 'stabilizing effect of intransitivities' \cite{allesina2011competitive,gilpin1975limit}.
The rock-paper-scissors (RPS) type game \cite{allesina2011competitive,gilpin1975limit,bhattacharyya2020mortality,szolnoki2014cyclic} is a classic example of intransitivity: rock beats scissor, scissor beats paper, but paper outperforms rock \cite{bhattacharyya2020mortality,szolnoki2014cyclic,reichenbach2007mobility,islam2022effect}. Coexistence is still possible when a large number of species participate in different variations of such game theoretical models  \cite{park2017emergence,allesina2011competitive,shi2010basins}.  Empirical evidence suggests that intransitive competition (RPS competition loop) appears to be common in bacterial strains \cite{kirkup2004antibiotic, kerr2002local}, phytoplankton communities \cite{soliveres2015intransitive}, Parasite-grass-forb \cite{cameron2009parasite}  or in general consumer-resource problems \cite{huisman1999biodiversity}. However, such a game theoretic approach does not always guarantee the stability of feasible equilibrium points \cite{gilpin1975limit,allesina2011competitive}, but rather provides a stable cycle in the majority of cases.
Weak perturbation, on the other hand, destabilizes the desired equilibrium points. However, theoretically one can show, beyond a pairwise interaction, higher order intransitive mixing in ecological communities can avoid such drawbacks by allowing the species to maintain stabilization \cite{grilli2017higher, battiston2020networks, letten2019mechanistic,mayfield2017higher,levine2017beyond,bairey2016high} by settling into a robust equilibrium point. A species is affected by the interaction of three or an even greater number of species in higher order
interactions  \cite{abrams1983arguments,lambiotte2019networks}. It should be noted that the role of intransitivity and intra as well as inter specific higher order interactions are tested in density dependent cyclic dynamics \cite{stouffer2018cyclic}, particularly in plant dynamics, using data from four co-occurring annual plant species in SW Western Australia. In contrast, habitat loss has the potential to reverse the effect of HOI \cite{li2020habitat}. Singh  {\it et al.\ } \cite{singh2021higher} demonstrated that the negative strength of interspecific HOI determines the fate of the  bio-diversity.

\par Here we revisit the classical higher order model developed by Grilli {\it et al} \cite{grilli2017higher}. The question here is, under what conditions can a species or a group of species gain or lose abundance strength (in comparison to others)? Our research provides a suitable and useful tweaking of the intransitive matrix to provide the species' target abundance states.
The abundance here denotes the stable equilibrium point.  In particular, we focus on the role of perturbation in the elements of the matrix $A$, 
which encodes information about interactions between two species. 
If the system has $N$  species/seedlings
in a forest, then
$A$ is an
$N \times N$ 
matrix, where $A_{ij}$ 
describes the probability of winning for the $i$-th species  over the $j$-th species, i.e., the ability to fill the gap in a canopy among two seedlings. 
Hence, 
$A_{ij}+A_{ji}=1$ $\forall$ $i,j$. 
Thus, the diagonal entries of $A$ will be $0.5$.
If all the elements of 
$A$
are equal, i.e., $A_{ij}=0.5$,
it represents the neutral model, where as
$0-1$ 
structure provides a {\it tournament} matrix 
and coexistence is possible in the presence of intransitive cycles of competitive dominance \cite{allesina2011competitive,grilli2017higher}. 
In this paper, we look at the effect of perturbation in a higher order interaction (HOI) where the intransitive cycle is captured by a circulant matrix
\begin{equation}
A=\rm{circ}\big(0.5, \alpha, 1-\alpha, \cdots, \cdots, \cdots).
%\begin{pmatrix}
%0.5 & \alpha & 1-\alpha & \cdots & \alpha & 1-\alpha \\
%1-\alpha & 0.5 & \alpha &  \cdots & 1-\alpha & \alpha \\
%\ddots & \ddots & \ddots & \ddots & \ddots & \ddots \\
%\alpha & 1-\alpha & \alpha & \cdots & 1-\alpha & 0.5
%\end{pmatrix}
\label{eq4}
\end{equation}

Here the circulant matrix generated by a single element 
$\alpha \in [0,1]$.
If the square matrix 
$A$'s order $N$ is odd, 
the last entry will be $1-\alpha$.
However, if $N$
is an even number, this last entry will be $\alpha$.
In the presence of HOI, for 
$N$  
species, this circulant matrix steers the whole  system  to a globally stable equilibrium point 
$\frac{1}{N}$ (Sec.\ \eqref{Sec.2}), where $N$ is odd. The choice of such a circulant matrix provides an utterly ecological community with equal strength of interaction among two species, similar up to the direction. Thus, initially, we do not yield any extra privilege to any species in the unperturbed system. Furthermore, we try to inspect the effect of feasible perturbation in a specific pair of species.

\begin{figure*}[!t]
	\centerline{\includegraphics[width=0.5\textwidth]{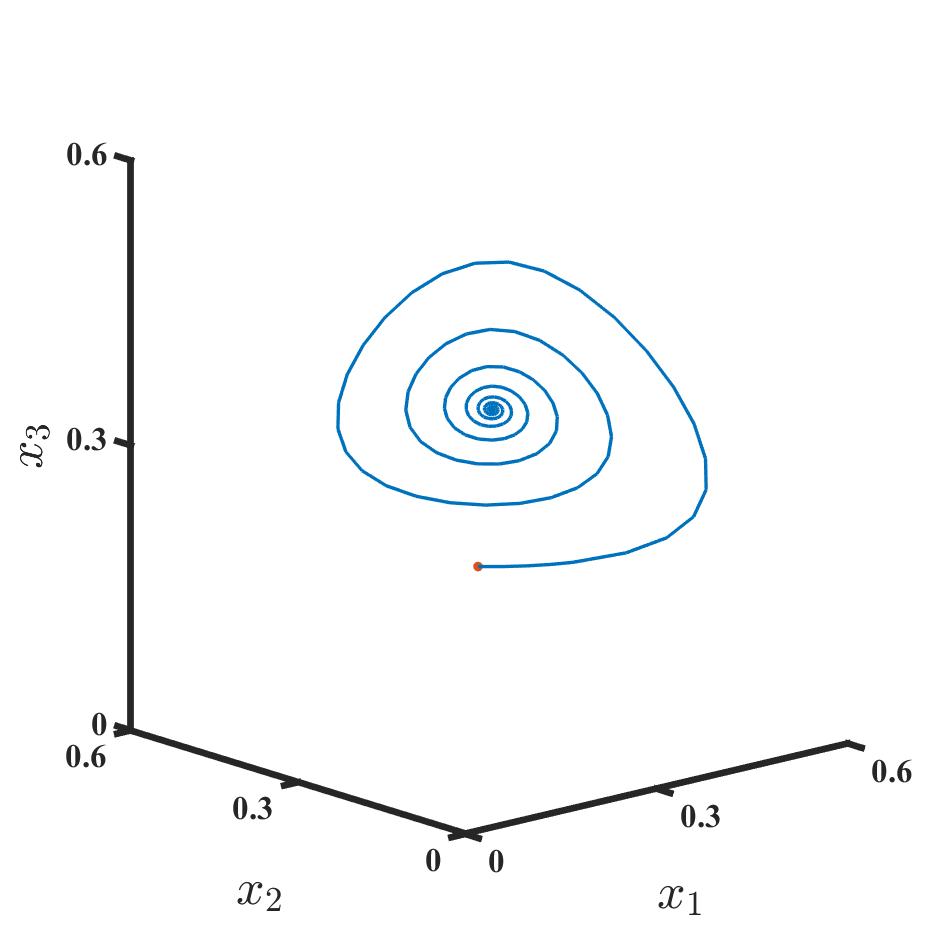}}
	\caption{{\bf Oscillation quenching and global convergence}: The system of differential equations \eqref{eq3} with the interaction matrix $A$ given in Eq.\ \eqref{eq4} converges to the stationary point $\bigg(\dfrac{1}{3},\dfrac{1}{3},\dfrac{1}{3}\bigg)$.
		The initial condition
		$(0.45,0.41,0.14)$ 
		is chosen randomly within the
		$(0,1) \times (0,1) \times (0,1)$ 
		maintaining the constraint 
		$x_1(0)+x_2(0)+x_3(0)=1.0$. Here, 
		$\alpha=0.75$.}
	\label{Picture1}
\end{figure*}

We investigate the effect of {\textit{localized perturbation}} on the identical equilibrium points. Here local perturbation signifies that if $\alpha$ value between two species are slightly changed, the rest of the interaction remains unchanged/unperturbed. We have shown analytically that the density of both perturbed species maintains a relation in the form of a rectangular hyperbola with respect to the applicable perturbation's strength for fixed $\alpha$. Thus, the respective densities of the perturbed species are either increased or decreased depending on the strength of the perturbation. 
Specifically, the species density of each of those perturbed species increases if their internal competition reduces, i.e., they try to reach neutral competition \big($||A_{ij}-A_{ji}|| \to 0$\big). In contrast, the density of both species decreases if they increase the strength of competition. A schematic figure is drawn to portray this scenario in Fig.\ \eqref{prac11} \, where in the subfigure (a), all species are unperturbed, maintaining the original strength following the interaction matrix $A$. The slight increment of winning probability of species 1 against species 2, keeping the other species unperturbed, leads to a reduction in the respective densities of the perturbed species (See Fig.\ \eqref{prac11} (b)). A reverse scenario of enhancing the winning probability of species 2 against species 1 is contemplated in Fig.\ \eqref{prac11} (c), where interestingly, the density of both species increases compared to the unperturbed species.

We have also found the two-clustered equilibrium points of perturbed 
(say $K$  species) 
and unperturbed 
($N-K$)
species analytically. As per our systematic approach, the perturbed species' density varies within the interval $[0.5,1]$. Similarly, by applying $w$ number of perturbations to $2w$ species among $N \geq (2w+1)$ species, we can split the unique equilibrium points into $(w+1)$ number of different clusters. Finally, we relax the perturbation scheme and analytically calculate the inhomogeneous stable equilibrium of $N=3$  species where all the elements in the upper triangular part of the circular matrix are perturbed within a specific range. Thus, we are able to demonstrate our findings in a general set-up, where the interaction matrix $A$ attains all intermediate values along with the neutral and tournament matrix. Interestingly, if more than two perturbations have been applied, then the relative change in density of one perturbed group relative to another is inversely proportional to its own perturbed strength. Note that perturbations can be considered evolutions of traits in our system of interacting species. In the presence of selection pressure, traits can be evolved in such a way that the traits achieve a significant advantage during any competition. %While discussing the types of perturbation, we can show how it is linked with the evolution of stronger or weaker species.
A similar analogy is already well-established in predator-prey systems \cite{abrams2000evolution}.

\section{Mathematical Model}

\par We consider a dynamical model describing the temporal evolution of the density $x_i(t)$ of the $i$-th species as follows  \cite{grilli2017higher}

\begin{equation} \label{eq1}
\begin{split}
{\dot{x}}_{i}=x_i \bigg[ \sum_{j=1}^{N} \sum_{k=1}^{N} \big( 2A_{ij}A_{ik}+A_{ij}A_{jk}+A_{ik}A_{kj}\big)x_{j} x_{k} -\eta_i \bigg], \\ i=1,2,\dots,N.
\end{split}
\end{equation}

Here, 
$\eta_i$
is the mortality rate of the 
$i$-th 
species. The death rate 
$\eta_i=\eta=1.0$
is chosen to be identical for all the species. Since, 
$x_i(t)$ 
represents the density of the $i$-th species, thus the sum of all species density is one, i.e., 
$\sum_{i=1}^{N} x_{i}(t)=1$.
%   $A$ is an $N \times N$ 
%  matrix, where $A_{ij}$ describes the probability of winning for the $i$-th species over the $j$-th species. Hence, 
% $A_{ij}+A_{ji}=1$ $\forall$ $i,j$. 
%   Thus, the diagonal entries of $A$ will be $0.5$. 
Instead of the pairwise interaction, we consider the interactions among the possible triplets, where the two species compete among each other and the winner play against the third one. The term within the bracket
$\big(2A_{ij}A_{ik}+A_{ij}A_{jk}+A_{ik}A_{kj}\big)$ 
portrays the winning probability of $i$-th species over other two $j$ and $k$ species under all possible circumstances. 
$A_{ij}A_{jk}$ 
describes the probability of winning of the $j$-th species over the $k$-th species and, ultimately, the $i$-th species dominate the $j$-th species. Similarly, the term 
$A_{ik}A_{kj}$ 
represents $k$-th species win over the $j$-th species, and then, the $i$-th species beat the $k$-th species. The probability of winning of $i$-th species over the other two $j$ and $k$ species is encapsulated by
$A_{ij}A_{ik}$. The factor $2$ here depicts that $i$-th species can play first with $j$-th species, and then with $k$-th species or vice-versa. Since, we choose $\eta_i=1$ and $\sum_{i=1}^{N} x_i(t)=1$, thus Eq.\ \eqref{eq1} transforms to

\begin{equation}\label{eq2}
{\dot{x}}_{i}=x_i \bigg[ \sum_{j=1}^{N} \sum_{k=1}^{N} \big( 2A_{ij}A_{ik}+A_{ij}A_{jk}+A_{ik}A_{kj}-1\big)x_{j} x_{k} \bigg].
\end{equation}

%Now,
%
%\begin{equation}
%\begin{aligned}
%\begin{split}
% 2A_{ij}A_{ik}+A_{ij}A_{jk}+A_{ik}A_{kj}-1\\
% =2A_{ij}A_{ik}+(1-A_{ji})A_{jk}+(1-A_{ki})A_{kj}-1 \\ =2A_{ij}A_{ik}-A_{ji}A_{jk}-A_{ki}A_{kj}=B_{ijk} \hspace{0.3cm} \text{(say)}.
% \end{split}
% \end{aligned}
% \label{eq38}
% \end{equation}

Then, Eq.\ \eqref{eq2} reduces to the form,

\begin{equation}\label{eq3}
{\dot{x}}_{i}=x_i \bigg[ \sum_{j=1}^{N} \sum_{k=1}^{N} B_{ijk} x_{j} x_{k} \bigg],
\end{equation}

where $B_{ijk}=2A_{ij}A_{ik}+A_{ij}A_{jk}+A_{ik}A_{kj}-1
=2A_{ij}A_{ik}-A_{ji}A_{jk}-A_{ki}A_{kj}$, since $A_{ij}+A_{ji}=1$ $\forall i,j=1,2,\dots,N$.

{\color{black} Note that there is a thin difference between our model and the RPS game. The classical RPS game is basically a three-player zero-sum game where cyclic dominance is observed. We consider a winning probability between two species as opposed to the classical RPS, wherein in a game, one species beats another with $100\%$ possibility. Moreover, our model describes the competition between $N \geq 3$ species rather than three species. One may quickly obtain a suitable variant of the RPS game by choosing suitable $\alpha$ and $N$.} %Note that this replicator equation describes a generalized version of the RPS game, essentially a three-player zero-sum game where cyclic dominance is observed. We consider a winning probability between two species as opposed to the classical RPS, wherein in a game, one species beats another with $100\%$ possibility. Moreover, our model describes the competition between $N \geq 3$ species rather than three species.}

\begin{figure*}[!t]
	\centerline{\includegraphics[width=1.0\textwidth]{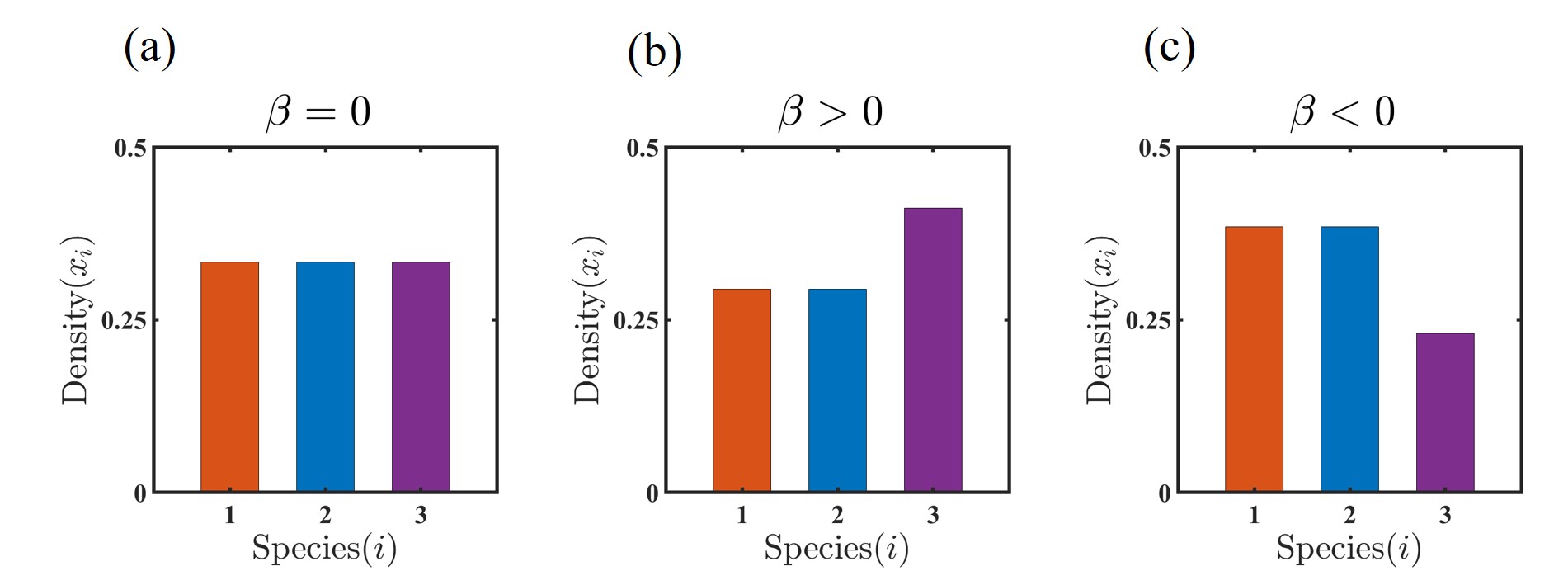}}
	\caption{{\bf Effect of localized perturbation}: The species densities vary for different choices of 
		$\beta$ for fixed $\alpha=0.75$. 
		(a) Initially, for
		$\beta=0$, 
		the unperturbed system leads to the coexistence of all species with equal densities. However, the scenario changes with the choices of non-zero $\beta$. (b) For our specific choice of 
		$\alpha > 0.5$, $\beta>0$ 
		reduces the first and second species' density (orange and blue respectively) and, consequently, raises $x_3$ (purple). 
		(c) $\beta<0$ improves (raises) 
		$x_1$ and $x_2$ 
		compared to the unperturbed case, and diminishes $x_3$. 
		The results are valid for any choice of initial conditions maintaining the relation 
		$x_1(0)+x_2(0)+x_3(0)=1$. 
		For further information, please see the main text Sec.\  4.\eqref{perb_N3}}.
	\label{Picture2}
\end{figure*}

%We restrict the choice of $A$ for our investigation by choosing a special kind of circulant matrix generated by a single element $\alpha \in [0,1]$ as follows,

%\begin{equation}
%A=circ\big(0.5, \alpha, 1-\alpha, \cdots, \cdots, \cdots).
%\begin{pmatrix}
%0.5 & \alpha & 1-\alpha & \cdots & \alpha & 1-\alpha \\
%1-\alpha & 0.5 & \alpha &  \cdots & 1-\alpha & \alpha \\
%\ddots & \ddots & \ddots & \ddots & \ddots & \ddots \\
%\alpha & 1-\alpha & \alpha & \cdots & 1-\alpha & 0.5
%\end{pmatrix}
%\label{eq4}
%\end{equation}

%If the order $N$ of the square matrix $A$ is odd, then the last entry of Eq.\ \eqref{eq4} will be $1-\alpha$. However, this last entry will be $\alpha$ if $N$ is even. 

\section{The dynamics of the replicator equations} \label{Sec.2}

\par To determine the optimal strategy for the matrix $A$ by adopting the methods provided in Ref.\ \cite{spaniel2014game}, we assume $p_i \in [0,1]$ is the probability of the $i$-th species with density $x_i$, $i=1,2,3,\cdots,N$. 
Let us further assume, 
$N \geq 3$ 
is odd. The existence of the interior stationary point with all non-zero components is unachievable for the even number of species \cite{fisher1995optimal}. The impossibility of maintenance of biodiversity with an even number of species is emphasized on the {\color{blue}Appendix \eqref{Even}}. Let 
$\mathbf{p}=\big(p_1,p_2,p_3,\cdots,p_N \big)^T$ 
be the stochastic vector with 
$0 \leq p_i \leq 1$. % $\forall$ $i$. 
The expected values of each species's density for the row player are 
\begin{equation}\label{eq5}
\begin{aligned}
\begin{split}
E(x_1)=(A\mathbf{p})_1=0.5p_1+\alpha(p_2+p_4+\cdots+p_{N-1})\\+(1-\alpha)(p_3+p_5+\cdots+p_N),\\ 
E(x_2)=(A\mathbf{p})_2=(1-\alpha)(p_1+p_4+p_6+\cdots+p_{N-1})\\+0.5p_2+\alpha(p_3+p_5+\cdots+p_N),\\
\vdots \hspace{5cm} \\
E(x_N)=(A\mathbf{p})_N=\alpha(p_1+p_3+\cdots+p_{N-2})\\+(1-\alpha)(p_2+p_4+\cdots+p_{N-1})+0.5p_N,
\end{split}
\end{aligned}
\end{equation}
along with $\sum_{i=1}^{N}p_i=1$. Here, $(A\mathbf{p})_i$ represents the $i$-th row of the matrix $A\mathbf{p}$. We mainly assume $\alpha \neq 0.5$ to avoid the occurrence of the rank-deficient matrix for which the matrix $A$ provides multiple optimal strategies. Solving Eq.\ \eqref{eq5}, we obtain $p_i=\dfrac{1}{N}$, $i=1,2,\dots,N$. Thus, the optimal strategy suggests all species will coexist with equal densities. Substituting $x_i=\dfrac{1}{N}$ in Eq.\ \eqref{eq3} for odd $N$, we find $\bigg(\dfrac{1}{N},\dfrac{1}{N},\cdots,\dfrac{1}{N}\bigg)$ is a stationary point of the Eq.\ \eqref{eq3}, as 

\begin{equation} \label{eq6}
\begin{aligned}
\sum_{j=1}^{N} \sum_{k=1}^{N} B_{ijk}=0, \hspace{0.5cm} i=1,2,\dots,N.
\end{aligned}
\end{equation}

Thus, the equilibrium $(p_1,p_2,\cdots,p_N)$ calculated through the interaction matrix $A$ serves as a stationary point $(x_1=p_1,x_2=p_2,\cdots,x_N=p_N)$ of the system \eqref{eq3}. For the validation, we first choose the simpliest case of $N=3$. We find the system converges to the stationary point $\bigg(\dfrac{1}{3},\dfrac{1}{3},\dfrac{1}{3}\bigg)$ 
for all initial conditions chosen from 
$(0,1) \times (0,1) \times (0,1)$.
The initial conditions are chosen maintaining the constraint $x_1(0)+x_2(0)+x_3(0)=1.0$. The local stability analysis provides the following three eigen values 
\begin{equation}\label{eq7}
\begin{aligned}
\lambda_{1,2}=-0.0625 \pm 0.4315i \text{and} %\\ 
\hspace{.1cm}
\lambda_3=-0.0207,
\end{aligned}
\end{equation}
and hence, it suggests the stationary point $\bigg(\dfrac{1}{3},\dfrac{1}{3},\dfrac{1}{3}\bigg)$ is a stable focus-node. In fact, we have performed a global stability analysis by choosing the following Lyapunov function,
\begin{equation}\label{eq8}
\begin{aligned}
V(x_i)=-\dfrac{1}{N} \sum_{i=1}^{N} \log(Nx_i),
\end{aligned}
\end{equation}  
where $N \geq 3$ 
is an odd integer. Since $\log$ is a concave function, thus by using Jensen's inequality \cite{jensen1906fonctions,dekking2005modern,ruel1999jensen}, we find
\begin{equation}\label{eq36}
\begin{aligned}
\sum_{i=1}^{N} \frac{1}{N} \log (Nx_i) 
\leq \log \bigg( \sum_{i=1}^{N} \frac{1}{N}  (Nx_i) \bigg)=\log  \sum_{i=1}^{N} x_i=\log (1)=0.
\end{aligned}
\end{equation}
Thus, clearly $V(x_i) \geq 0$, for $x_i \in (0,1)$ and %$V\bigg(\dfrac{1}{N}\bigg)=0$. 
\begin{equation}\label{eq37}
V\bigg(\dfrac{1}{N}\bigg)=-\dfrac{1}{N} \sum_{i=1}^{N} \log \bigg(\frac{N}{N} \bigg)=0.
\end{equation}

Let $\xi_j$ be a non-zero perturbation such that 

\begin{equation}\label{eq9}
\xi_j=x_j-\dfrac{1}{N}, j=1,2,\cdots,N.
\end{equation} 

Then, $\sum_{j=1}^{N} \xi_j=0$. Now,

\begin{equation*}
%\begin{aligned}
\begin{aligned}
\begin{split}
\frac{dV}{dt} %= -\frac{1}{N} \sum_{i=1}^{N} \frac{1}{x_i} \frac{dx_i}{dt} %\\[\parskip]
= -\frac{2}{N} \sum_{i=1}^{N} \sum_{j=1}^{N} \sum_{k=1}^{N} \big(A_{ij}A_{ik}+A_{ij}A_{jk}\big)x_jx_k + 1\\[\parskip]
=-2 \sum_{j=1}^{N} \sum_{k=1}^{N} \big(\sum_{i=1}^{N} \frac{1}{N} A_{ij} \big)A_{jk}x_jx_k \\-\frac{2}{N} \sum_{i=1}^{N} \big(\sum_{j=1}^{N} A_{ij} x_j\big) \big(\sum_{k=1}^{N} A_{ik} x_k\big) + 1 \\[\parskip]
%=- \sum_{j=1}^{N} \sum_{k=1}^{N} A_{jk}x_jx_k-\frac{2}{N} \sum_{i=1}^{N} \big(\sum_{j=1}^{N} A_{ij} x_j\big)^2+1, \text{since} \sum_{i=1}^{N} \frac{1}{N} A_{ij}= \dfrac{\frac{N-1}{2}+0.5}{N} =\frac{1}{2} \\[\parskip]
%=-\frac{2}{N} \sum_{i=1}^{N} \big(\sum_{j=1}^{N} A_{ij} x_j\big)^2+\frac{1}{2}, \text{since} \sum_{j=1}^{N} \sum_{k=1}^{N} A_{jk}x_jx_k=\frac{1}{2}\\[\parskip]
\end{split}
\end{aligned}
\end{equation*}

Since $\sum_{i=1}^{N} \frac{1}{N} A_{ij}= \dfrac{\frac{N-1}{2}+0.5}{N} =\frac{1}{2}$ and $\sum_{j=1}^{N} \sum_{k=1}^{N} A_{jk}x_jx_k=\frac{1}{2}$, we have

\begin{equation*}
%\begin{aligned}
\begin{aligned}
\begin{split}
\frac{dV}{dt}
=-\frac{2}{N} \sum_{i=1}^{N} \big(\sum_{j=1}^{N} A_{ij} x_j\big)^2+\frac{1}{2}\\
=-\frac{2}{N} \sum_{i=1}^{N} \bigg(\sum_{j=1}^{N} A_{ij} \bigg(\frac{1}{N}+\xi_j\bigg)\bigg)^2+\frac{1}{2}\\[\parskip]
%=-\frac{2}{N} \sum_{i=1}^{N} \bigg(\frac{1}{2}+\sum_{j=1}^{N} A_{ij} \xi_j\bigg)^2+\frac{1}{2}\\[\parskip]
=-\frac{2}{N} \sum_{i=1}^{N} \bigg(\sum_{j=1}^{N} A_{ij} \xi_j\bigg)^2-\frac{2}{N} \sum_{i=1}^{N} \sum_{j=1}^{N} A_{ij} \xi_j \\ %\\[\parskip]
%=-\frac{2}{N} \sum_{i=1}^{N} \bigg(\sum_{j=1}^{N} A_{ij} \xi_j\bigg)^2-2 \sum_{j=1}^{N} \sum_{i=1}^{N} \bigg(\frac{1}{N}A_{ij}\bigg) \xi_j \\[\parskip]
=-\frac{2}{N} \sum_{i=1}^{N} \bigg(\sum_{j=1}^{N} A_{ij} \xi_j\bigg)^2- \sum_{j=1}^{N} \xi_j \\[\parskip]
=-\frac{2}{N} \sum_{i=1}^{N} \bigg(\sum_{j=1}^{N} A_{ij} \xi_j\bigg)^2 < 0,
%\end{aligned}
\end{split}
\end{aligned}
\end{equation*}

unless, $A$ is a rank deficient matrix. Thus, the stationary point $\bigg(\dfrac{1}{N},\dfrac{1}{N},\cdots,\dfrac{1}{N}\bigg)$ is globally stable with odd $N$. Therefore, the system \eqref{eq3} may possess several stationary points; nevertheless, the system always converges to the globally stable stationary point predicted by the equilibrium calculated through the interaction matrix $A$ given in \eqref{eq4} for odd $N \in \mathbb{N}$.

\section{Consequences of perturbation among the interacting species}
\label{Local-perb}
\subsection{Perturbation in single species-pair $(N=3)$}
\label{perb_N3}
\par Now, we want to understand the impact of a perturbation at a single entry of the interaction matrix $A$. \textit{Does it alter the globally stable stationary point? If yes, can we predict it before investigating the dynamics through numerical simulations?} Again, we consider the simplified choice to investigate this and begin with $N=3$. The perturbed matrix looks like

\begin{equation}\label{eq10}
\Tilde{A} = 
\begin{bmatrix}
0.5 & \alpha + \beta & 1-\alpha \\
1-\alpha -\beta & 0.5 & \alpha \\
\alpha & 1-\alpha & 0.5 
\end{bmatrix}.
\end{equation}

%To distinguish with the unperturbed matrix $A$, we have labeled this new matrix as $\Tilde{A}$. 
Still this new interaction matrix satisfies 
$\Tilde{A}_{ij} + \Tilde{A}_{ji} = 1$, 
as each entry here too contemplates the probability of winning among the species. Note that we just perturb a single pair of species' interaction, which is 
$A_{12}$ and consequently $A_{21}$ 
without loss of any generality in this case. Clearly, 
$\beta$ may be positive or negative, however we choose the specific interval which can mimick the  physically meaningful interaction matrix. The choice of $\beta$ is based on the following two criteria, %$\beta$ is chosen mainly maintaining the following two criteria,

\begin{itemize}
	\item One needs to choose $\beta$ in such a way that each entry of the interaction matrix lies within $[0,1]$ and indicates a non-negative real number representing a probability of winning the pairwise competition.
	
	\item Applying $\beta$ does not change the role of stronger and weaker species; it only increases or reduces the interaction between two competitors. This is to make sure that every species coexist after applying perturbation. Otherwise, before perturbation, every species was stronger and weaker in the same number of interactions; reversing the role will form a disparity and eventually lead to the destruction of all species' coexistence.
\end{itemize}

\begin{figure*}[!t]
	\centerline{\includegraphics[width=1.0\textwidth]{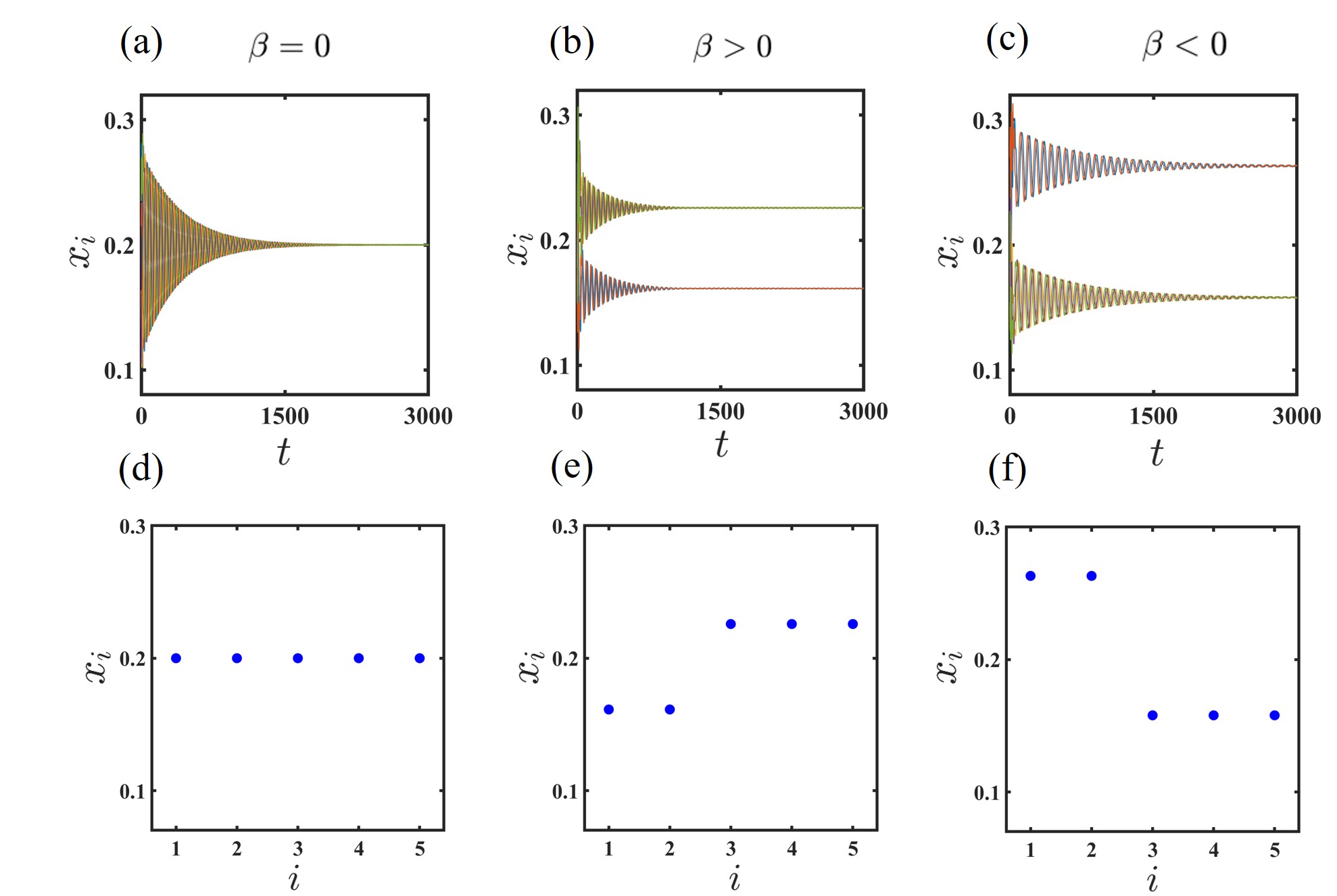}}
	\caption{{\bf Impact of a single perturbation on the first-second species interaction}: (a, d) A common convergence for all the species in an identical density is observed for the unperturbed case with $\beta=0$. (b, e) For $\beta=0.1$, the external perturbation reduces their respective densities of first and second species. The emergence of two clusters is represented in subfigure (e) after the initial transient. (c, f) For $\beta=-0.1$, the internal perturbation enhances the densities of two perturbed species. The unperturbed three species maintain a coherent state. All these observations remain valid for any permissible choice of initial conditions. Here, $\alpha=0.75$ and $N=5$.}
	\label{Picture4}
\end{figure*}

Here 
$\beta=0$ leads to $A=\Tilde{A}$. 
Irrespective of the sign of 
$\beta$,
one can determine the optimal strategy by equating the expected values of all three species corresponding to the row. We have the following three equations

\begin{equation}\label{eq11}
\begin{aligned}
E(x_{1}) = (1-\alpha) + (\alpha -0.5)p_{1} + (2\alpha + \beta -1)p_{2},\\
E(x_{2}) = \alpha + (1 - 2\alpha - \beta)p_{1} + (0.5 - \alpha)p_{2},\\
E(x_{3}) = 0.5 + (\alpha -0.5)p_{1} + (0.5 - \alpha)p_{2},
\end{aligned}
\end{equation}

where 
$p_1$, $p_2$, and $p_3=1-p_1-p_2$ 
are the probabilities of species survivability. Solving the Eq.\ \eqref{eq11}, we obtain 

\begin{equation}\label{eq12}
\begin{aligned}
p_{1} = \frac{\alpha - 0.5}{3(\alpha - 0.5) + \beta},\\ 
p_{2} = \frac{\alpha - 0.5}{3(\alpha - 0.5) + \beta},\\
p_3=\frac{(\alpha - 0.5)+\beta}{3(\alpha - 0.5) + \beta}.
\end{aligned}
\end{equation}

\begin{figure*}[!t]
	\centerline{\includegraphics[width=1.0\textwidth]{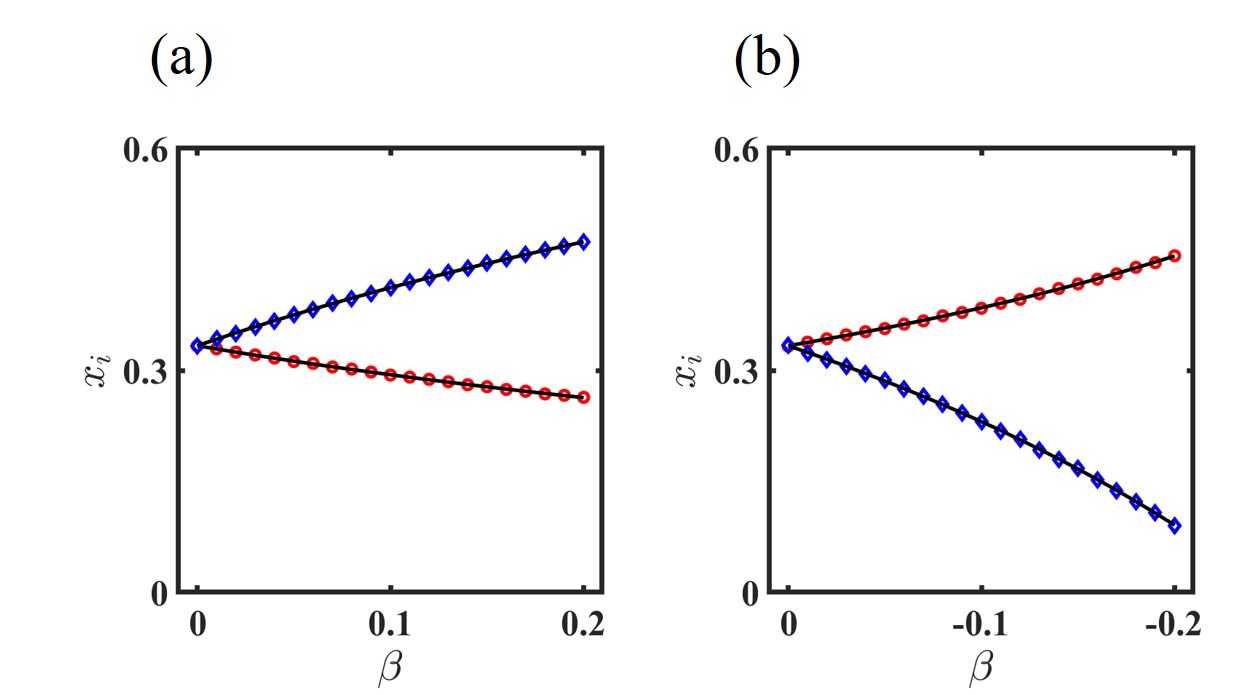}}
	\caption{{\bf The impact of localized perturbation on a single pair of interaction}: We have perturbed the first and second species' density by $\beta \neq 0$.
		Due to our choice of 
		$\alpha=0.75>0.50$, $\beta$ can be varied within 
		$[-0.25,0.25]$.
		However, we ignore the case for 
		$\beta=-0.25$ 
		to avoid the rank deficient matrix 
		$\Tilde{A}$. 
		If we choose 
		$\beta<-0.25$, 
		the interaction between the first and the second species gets reversed, so we disregard these values of 
		$\beta$. For 
		$\beta > 0.25$, 
		$\alpha$ 
		exceeds the value $1$. (a) External perturbation ($\beta>0$).  
		Here the unperturbed species is marked by blue diamonds and first two perturbed species  are marked by red circles.  Here the  abundance of unperturbed   species increases slowly and the abundances of perturbed species are decreased slowly.   (b) Internal perturbation ($\beta<0$).  
		Here the abundance of unperturbed species (blue diamonds) is dropped rapidly. The analytical calculation fits perfectly with the numerically simulated values shown by the markers. }
	\label{Picture3}
\end{figure*}

Initially, we start with 
$\beta>0$ for 
$\alpha > 0.5$ and $\beta<0$
for $\alpha < 0.5$. 
The valid range for these choices of 
$\beta$ is $[-\kappa, 0)$ or $(0,-\kappa] $, 
where, $\kappa = {\rm min} \{ \alpha, 1-\alpha \}$.
For these choices of 
$\beta$, we have 
$||A_{ij}-A_{ji}|| < ||\Tilde{A}_{ij} - \Tilde{A}_{ji}||$. 
We refer to this state as the \textit{external perturbation} because this choice of
$\beta$ 
enhances the surviving probability of the $i$-th species (which is the first species in the case of the matrix \eqref{eq10}) when interacted with the $j$-th species (the second species for our particular choice in matrix \eqref{eq10}). To scrutinize the impact of 
$\beta$, 
we choose a particular value of 
$\beta=0.1$ and $\alpha=0.75$. 
The numerical integration of Eq.\ \eqref{eq3} with random permissible initial conditions provides 
$x_1=0.29412,x_2=0.29412,x_3=0.41176$
(See Fig.\ \eqref{Picture2} (b), orange, blue and purple respectively). It is a stable focus-node as the eigen values of the corresponding Jacobian are  

\begin{equation}\label{eq33}
\lambda_{1,2}=-0.0788 \pm 0.4792i \hspace{0.2cm} \text{and} \hspace{0.2cm} \lambda_3=-0.0226.
\end{equation}

%$\lambda_1, \lambda_2=-0.0788 \pm 0.4792i$ and $\lambda_3=-0.0226$.
Again, we need to choose the initial conditions maintaining the constraint 
$x_1(0)+x_2(0)+x_3(0)=1$. 
Interestingly, our numerically simulated values, in this case too, are well fitted with our analytically predicted values $\eqref{eq12}$. The densities of the first (orange) and second (blue) species reduce compared to the unperturbed 
($\beta=0$) 
case (See Fig.\ \eqref{Picture2} (a-b)). This scenario suggests external perturbation is helpful in gaining species density for the unperturbed species as $x_3>x_2=x_1$.
Apart from the analytical understanding, we realize this finding is absolutely recognizable through the physical interpretation. An external perturbation in the first-second interaction increases the probability of the survivability of the first species against the second species. This, in turn, decreases the density of the second species, and as the second species dominates the third species, the density of the third species goes up. On the other hand, as the third species dominate the first species, increasing $x_3$ implies firm control over the first species, lowering its density.
\par For $\beta < 0$ for
$\alpha>0.5$ and $\beta > 0$ for $\alpha<0.5$, 
we can define an {\it internal perturbation} for which we have 
$||A_{ij}-A_{ji}|| > ||\Tilde{A}_{ij} - \Tilde{A}_{ji}||$. 
The choice of 
$\beta$ 
is restricted within the range 
$(-\rho, 0)$ or $(0,\rho)$, 
where $\rho = \frac{1}{2} |1 - 2\alpha|$. 
{We call this as internal perturbation as it is reducing the competition between them leading them to the equal cooperation. }
We have encapsulated all the requisite information regarding the classification of internal and external perturbation in Table \eqref{tab:multicol}. However, we discard the case of 
$\alpha=0.5$, 
which creates a rank deficient matrix that yields multistable system \eqref{eq3}. We choose 
$\beta=-0.1$ for $\alpha=0.75$, 
we find the densities of the first and second species increase (orange and blue respectively) and the third species' density (purple) decreases (See Fig.\ \eqref{Picture2} (c)). An internal perturbation in the first-second species interaction decreases the surviving probability of the first species against the second species. This, in turn, increases the density of the second species, and as the second species dominates the third species, the density of the third species goes down. On the other hand, as the third species dominate the first species, a decrease in $x_3$ implies weak control over the first species, raising its density. Moreover, the simulated result in Fig.\ \eqref{Picture2} (c) yields $x_1=x_2=0.38462,x_3=0.23077$, which fits with our derived equilibrium \eqref{eq12} of $\Tilde{A}$ for $\alpha=0.75$ and $\beta=-0.1$. This stationary point is a focus-node as per the local stability analysis with

\begin{equation}\label{eq34}
\lambda_{1,2}=-0.0422 \pm 0.3594i \hspace{0.2cm} \text{and} \hspace{0.2cm} \lambda_3=-0.0166.
\end{equation}

\begin{table}[ht]
	\caption{Distinguishing between internal and external perturbation depending on the choice of $\alpha$ and the permissible range of $\beta$ applied on the first and second species' interaction}
	\begin{center}
		\begin{tabular}{|c|c|c|}
			\hline
			\multirow{2}{*}{ $\alpha < 0.5$}& {\cellcolor{white} $\beta >0$}
			& {\cellcolor{white} Internal perturbation}\\  \cline{2-3}
			& {\cellcolor{white} $\beta < 0$}  & {\cellcolor{white} External perturbation}\\
			\hline
			$\alpha = 0.5$ & {\cellcolor{white} $\beta \neq 0$} & {\cellcolor{white} External perturbation}\\
			\hline
			\multirow{2}{*}{$\alpha > 0.5$}& {\cellcolor{white} $\beta >0$}
			& {\cellcolor{white} External perturbation}\\ \cline{2-3}
			& {\cellcolor{white} $\beta < 0$}  & {\cellcolor{white} Internal perturbation}\\  
			\hline
		\end{tabular}
	\end{center}
	\label{tab:multicol}
\end{table}

%$\lambda_1,\lambda_2=-0.0422 \pm 0.3594i$, and $\lambda_3=-0.0166$.
Besides, its stability can also be confirmed by the similar global stability analysis provided in Sec.\ \eqref{Sec.2}. To prove global stability, one needs to choose the positive-definite Lyapunov function \cite{sodhan2021metapopulation} 

\begin{equation}\label{eq35}
V(x_i)=- \sum_{i=1}^{N} {x}^*_i \log{\frac{x_i}{{x}^*_i}},
\end{equation}

where ${x}^*_i \in (0,1)$ is the $i$-th component of the stationary point of the system \eqref{eq3} with odd $N \in \mathbb{N}$, and proceeds like our previous analysis to show the time derivative of this function is globally negative definite.

\begin{figure*}[!t]
	\centerline{\includegraphics[width=0.85\textwidth]{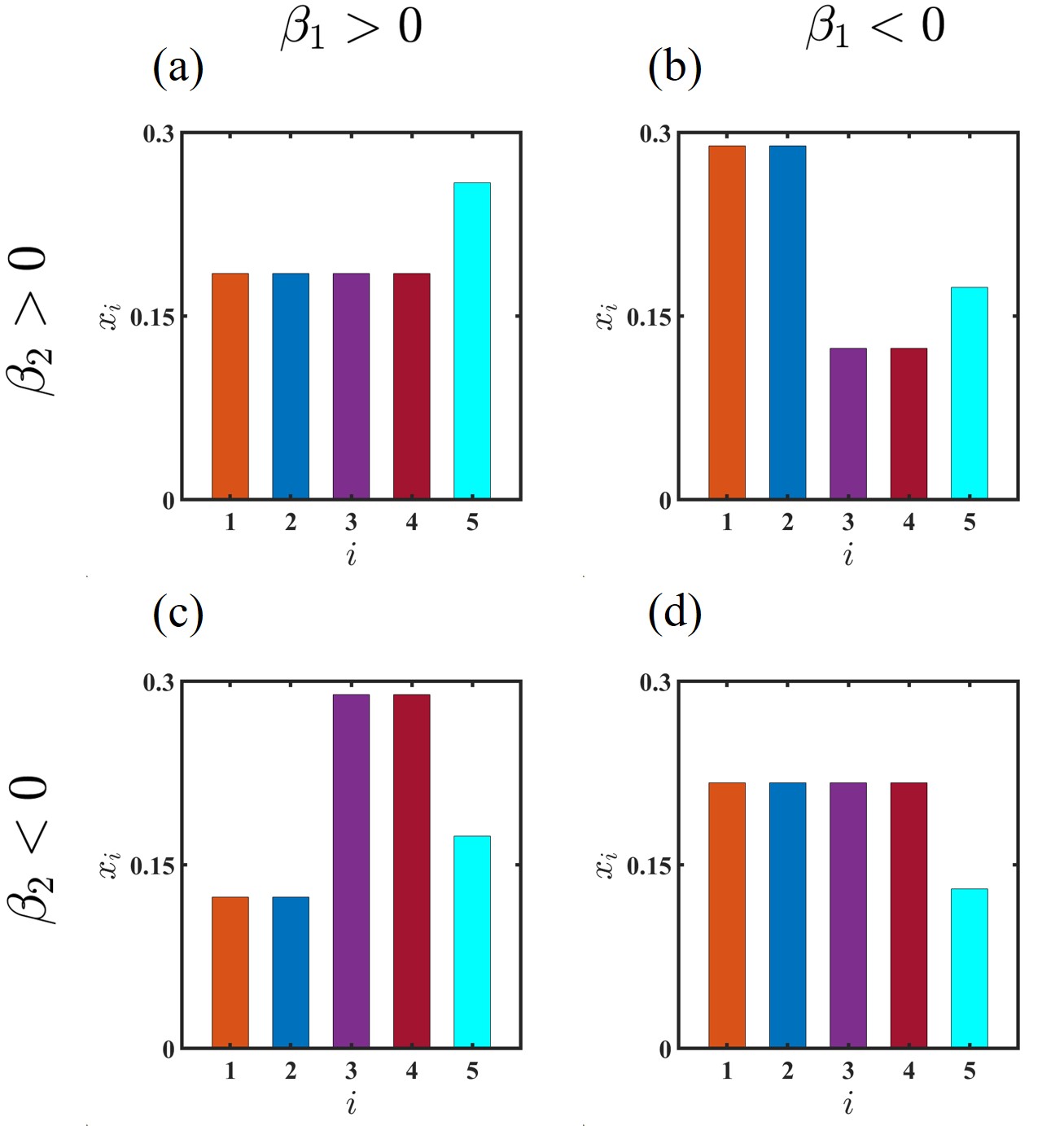}}
	\caption{{\bf Repercussion of two perturbations}: For $\alpha=0.75$, we represent four different cases. (a) $\beta_1=\beta_2=0.1$ 
		provides an improved density for the fifth species, while the rest of the densities decline compared to 
		$\beta_1=\beta_2=0$.
		(b) Through the consideration of $-\beta_1=\beta_2=0.1$, 
		we introduce the interplay of internal and external perturbations on the $N=5$ species. (c) Another different combination of external and internal perturbation with
		$\beta_1=-\beta_2=0.1$
		demonstrate the variation of species densities, where $x_1=x_2$ and $x_3=x_4$.
		(d) $\beta_1=\beta_2=-0.1$ 
		displays the influence of two internal perturbations on the first two species. Simulations are done with random permissible set of initial conditions. Initially for $\beta_1=\beta_2=0$, all species densities $x_1=x_2=x_3=x_4=x_5=0.2$ are equal. For further details, please see the main text.
	}
	\label{Picture5}
\end{figure*}

\par We confirm our claim of predicting the asymptotic values of each species' densities using Eq.\ \eqref{eq12} without solving it (Eq.\ \eqref{eq3}) numerically. By keeping fixed $\alpha=0.75$, we vary $\beta$ within the interval 
$[-0.2,0]$ 
for the internal perturbation and within 
$[0,0.2]$ 
for the external perturbation. The employed external perturbation 
($\beta > 0$) with 
$\alpha=0.75$ 
destroys the coherence among each trajectory in the asymptotic state. Since we have applied the perturbation only in the first and second species' density, those two trajectories split from the synchronous state. Satisfying the calculated stationary points in \eqref{eq12} shown by solid lines in Fig.\ \eqref{Picture3} (a), we find the species density of the first and second species diminish (red circles) with an increment of 
$\beta \in [0,0.2]$, 
and as a result of that, the density of third species (blue diamonds) increases. Our analytical prediction through the equilibrium \eqref{eq12} of 
$\Tilde{A}$ 
suggests
$x_1=x_2$ 
(red circles) 
increase  and 
$x_3$ 
(blue diamonds)
diminishes for negative 
$\beta \in [-0.2,0]$ 
with $\alpha=0.75$. 
Even the numerically simulated values shown by the marker in Fig.\ \eqref{Picture3} (b) perfectly match the analytical predictions. 
Thus a suitable  single perturbation impacting over a two-species interaction  can control the density of each species. Particularly, partially motivated by the analysis of Connolly {\it et al.\ } \cite{connolly2001interspecific}  we can measure the relative steady abundance between the perturbed and unperturbed species as follows (using Eq.\ \eqref{eq12})

\begin{equation} \label{eq 450}
\Delta_R=\frac{p_{\rm perb}}{p_{\rm unperb }}=\frac{\alpha-0.5}{(\alpha-0.5)+\beta}.
\end{equation}

\begin{figure*}[!t]
	\centerline{\includegraphics[width=1.0\textwidth]{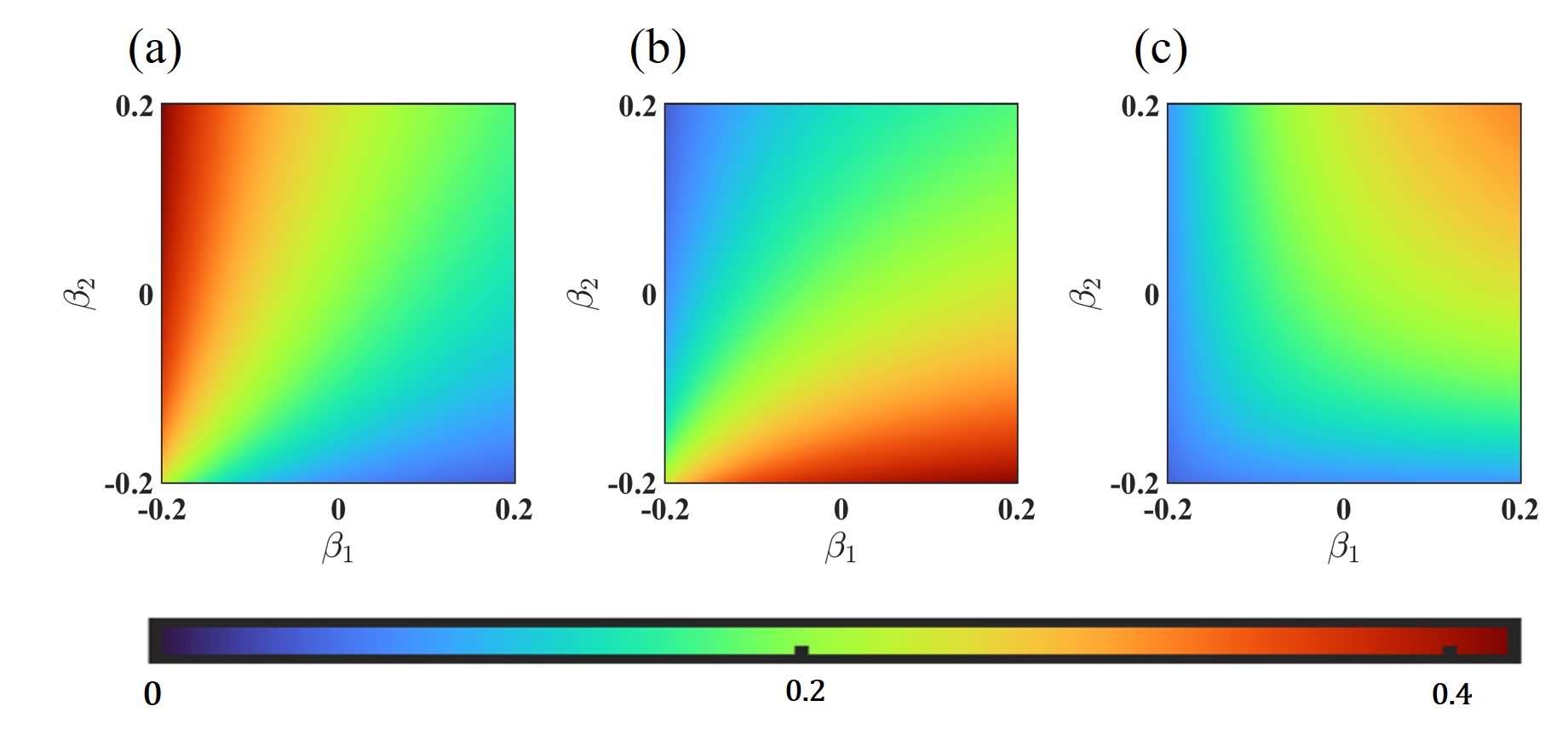}}
	\caption{{\bf Effect of two perturbations on the first four out of five species}: The triadic interaction-induced system diversities, i.e., non-trivial steady states, are explored using numerical methods (ODE45, Matlab R2021a). The choice of 
		$\beta_1 \neq \beta_2$ allows three distinct groups, viz.\ $x_1=x_2$, $x_3=x_4$, 
		and $x_5$. 
		We plot three different parameter space
		$\beta_1-\beta_2$ 
		with respect to 
		(a) $x_1(=x_2)$, (b) $x_3(=x_4)$, 
		and (c) $x_5$, 
		respectively for fixed $\alpha=0.75$. 
		The numerical simulations completely agree with our derived equilibrium \eqref{eq17} of $\Tilde{A}$ (See the matrix \eqref{eq16}). The system \eqref{eq3} is iterated for a sufficiently long number of iterations to disregard the initial transient and the final densities ($x_i$) are collected from the asymptotic behavior of the long term stationary state. Furthermore, we repeat the process for $50$ times independently with random permissible initial conditions to establish the generic behavior of the triple-wise competitive interaction and eliminate the initial conditions dependent finding. 
	}
	\label{Picture6}
\end{figure*}

For a suitable 
$\alpha$,  
the ratio between the unperturbed and perturbed species is inversely proportional to
$\beta$,
i.e., if 
$\beta$ 
increases in negative direction this ratio \eqref{eq 450} sharply increases  where as it slowly decays if 
$\beta$ 
increases in positive direction. This is also clear from the Fig.\ \eqref{Picture3} (a), where, in presence of positive 
$\beta$ 
the unperturbed species (red diamonds) slowly grows compared to perturbed species. On the contrary, it drops rapidly for negative 
$\beta$.
Thus, negative perturbation, i.e., a less competition helps the species to win strongly over the unperturbed species. %On the other hand, positive perturbation i.e., enhanced competition reduces the abundance of perturbed species, however the decay rate of losing abundance is slower compared to the decay rate of perturbed species when $\beta<0$.
In the next section we have generalized the perturbed scheme for 
$N$ species. 

\subsection{Perturbation in single species-pair (for any odd $N$)}

\par Again, we only consider an odd number of species as a comprehensive analysis is provided in {\color{blue}Appendix \eqref{Even}} to demonstrate the impossibility of the existence of the interior stationary point with an even number of species. Till now, although we have represented all the numerical simulations only for $N=3$. Nevertheless, one can easily derive the equilibrium of $\Tilde{A}$ for any odd $N \geq 3$ as follows,

\begin{equation}\label{eq13}
\begin{aligned}
x_{1} = x_{2} = \frac{(\alpha - 0.5)}{N(\alpha - 0.5) + (N-2)\beta}, \text{and} \hspace{0.1cm}  \\
x_{i} = \frac{(\alpha - 0.5)+\beta}{N(\alpha - 0.5) + (N-2)\beta}, \hspace{0.2cm}  \text{when} \hspace{0.2cm} i=3,4,\cdots,N.
\end{aligned}
\end{equation}

where the perturbation
$\beta \neq 0$, 
chosen from the permissible range, is applied only at 
${\Tilde{A}}_{12}$ 
and ${\Tilde{A}}_{21}$. We calculate this stationary point by following the similar analysis provided in Sec.\ \eqref{Sec.2}. We construct $N$ linear equations using the rows of $\Tilde{A}$, and solving those $N$ hyperplanes; we obtain Eq.\ \eqref{eq13}. To check the analytical expression of Eq.\ \eqref{eq13}, we consider here 5 species intransitive cycle  where the perturbation 
$\beta \neq 0$ 
is employed
at the interaction of the first and second species. The corresponding matrix looks like
\begin{equation}\label{eq14}
\Tilde{A} = 
\begin{bmatrix*}[r]
\phantom{-}0.5 & \alpha + \beta & 1-\alpha & \alpha & 1-\alpha \\
1-\alpha - \beta & 0.5 & \alpha & 1-\alpha & \alpha \\
\alpha & 1-\alpha & 0.5 & \alpha & 1-\alpha \\
1-\alpha & \alpha & 1-\alpha & 0.5 & \alpha \\
\alpha & 1-\alpha & \alpha & 1-\alpha & \phantom{-}0.5
\end{bmatrix*}.
\end{equation}
When $\beta=0$ with $\alpha=0.75$, 
the system \eqref{eq3} stabilizes to the stationary point $(0.2,0.2,0.2,0.2,0.2)$ 
coinciding with the derived equilibrium of 
$A$ for odd $N$, 
as portrayed in Fig.\ \eqref{Picture4} (a). Figure \eqref{Picture4} (d) depicts the snapshot at a particular time  after the transients. However, this scenario alters for non-zero 
$\beta$. For $\beta=0.1$ and $\alpha=0.75$, $x_1$ and $x_2$ 
split from the coherent group and we find 
$x_1=x_2=0.16129<x_i=0.22581$, for $i=3,4,5$. 
The asymptotic values of these two coherent groups fit with our derived values $\big(\zeta_1, \zeta_1, \zeta_2, \zeta_2, \zeta_2)$ at Eq.\ \eqref{eq13}, where $\zeta_1=\frac{(\alpha - 0.5)}{5(\alpha - 0.5) + 3\beta}$ and $\zeta_2=\frac{(\alpha - 0.5) + \beta}{5(\alpha - 0.5) + 3\beta}$.
%$\bigg(\frac{(\alpha - 0.5)}{5(\alpha - 0.5) + 3\beta}, \frac{(\alpha - 0.5)}{5(\alpha - 0.5) + 3\beta}, \frac{(\alpha - 0.5) + \beta}{5(\alpha - 0.5) + 3\beta}, \frac{(\alpha - 0.5) + \beta}{5(\alpha - 0.5) + 3\beta}, \frac{(\alpha - 0.5) + \beta}{5(\alpha - 0.5) + 3\beta}\bigg)$ at Eq.\ \eqref{eq13}. 
We plot the temporal evolution of each 
$x_i$ for $i=1,2,3,4,5$ 
in Fig.\ \eqref{Picture4} (b), and the snapshot at a particular time after the transient in Fig.\ \eqref{Picture4} (e). Apart from the external perturbation, we also examine the impact of internal perturbation on the interaction among the first and second species in subfigures (c) and (f) of Fig.\ \eqref{Picture4}. The choice of 
$\beta=-0.1$ 
leads to an increment of 
$x_1=x_2=0.26316$ 
as predicted by Eq.\ \eqref{eq13} compared to the unperturbed ($\beta=0$) scenario for 
$\alpha=0.75$. 
The value of 
$x_3=x_4=x_5=0.15789$ 
also satisfies our theoratical calculated value in Eq.\ \eqref{eq13}. So, applying the external perturbation at the first and second species’ interaction will reduce the corresponding densities. Whereas the internal perturbation always boosts the densities of the related species. The behavior of relative steady abundance $\Delta_R$ also follows the same trend. A relevant question we also raise here is what will happen if unequal perturbations are used for each pair of the species? We will deal with this concern in the following subsection. In the following subsection, we will investigate the influence of perturbation in multiple species.

%So, \st{the takeaway message} {\color{black} whenever we apply the internal perturbation at the first and second species' interaction, it will reduce the corresponding densities. Whereas the external perturbation always boosts the densities of the related species: {\color{magenta} STATEMENT IS CONFUSING.} The behaviour of relative steady abundance $\Delta_R$ also follows the same trend. A relevant question we may raise here, what will happen if unequal  perturbations are used for each of the pair of the species? }

\subsection{Perturbation in multiple species-pair (for any odd $N$)}
\label{perb_mutispecies}
\par Still, now, we have discussed the results of a single perturbation on the ensemble. However, one may introduce more than a single perturbation. We assume that a single species cannot be perturbed twice to investigate this scenario. By allowing 
$k$ 
perturbations, one can disturb the first 
$2k$ species. Thus, we strictly need 
$N>2k$. 
As $N$ is odd, we consider the greater than sign. The theoretical calculation yields the following equations predicting the asymptotic species densities of 
$N$ 
species as follows,

\begin{equation}\label{eq15}
\begin{aligned}
x_{i} = \frac{(\alpha - 0.5)}{N(\alpha - 0.5) + (N-2k)\beta}, \text{when} \hspace{0.2cm} i = 1,2,\cdots,2k \\ 
%\hspace{0.2cm} \text{and} \\
x_{i} = \frac{(\alpha - 0.5)+\beta}{N(\alpha - 0.5) + (N-2k)\beta}, \text{when} \hspace{0.2cm} i =2k+1,2k+2,\cdots,N.
\end{aligned}
\end{equation}

We find this solution by solving $N$ equations as shown in detail in Sec.\ \eqref{Sec.2}. The denominator contains the term $(N-2k)\beta$. This term appears as $k$ perturbations affect $2k$ species, and thus, the remaining $(N-2k)$ species will experience the influence of $\beta$. Thus perturbation in a single pair or identical perturbation in multiple species-species interaction split the system into two clusters steady states, and 
$\Delta_R$
follows the same trend as discussed before. Equations \eqref{eq15} easily leads to the earlier discussed results in Eq.\ \eqref{eq13} by setting $k=1$.

\par Thus our recipe provides a way to split the entire network into two clusters, and abundance of perturbed species pair will be increased if the pair decreases their competition.  Against this backdrop one may wonder whether this approach can increase the number of steady clusters and can control the abundance of each pair in the network.
The simple way is to introduce the different amounts of perturbation without disturbing the same species twice. For instance, let us consider the following interaction matrix for $N=5$

\begin{equation}\label{eq16}
\Tilde{A} = 
\begin{bmatrix}
0.5 & \alpha + \beta_{1} & 1-\alpha & \alpha & 1-\alpha \\
1-\alpha - \beta_{1} & 0.5 & \alpha & 1-\alpha & \alpha \\
\alpha & 1-\alpha & 0.5 & \alpha + \beta_{2} & 1-\alpha \\
1-\alpha & \alpha & 1-\alpha - \beta_{2} & 0.5 & \alpha \\
\alpha & 1-\alpha & \alpha & 1-\alpha & 0.5
\end{bmatrix}.
\end{equation}

To calculate the equilibrium of this matrix \eqref{eq16}, we equate the following five expressions,

\begin{equation}\label{eq5123}
\begin{aligned}
E(x_1)=(A\mathbf{x})_1=0.5x_1+(\alpha + \beta_{1})x_2+(1-\alpha)x_3+\alpha x_4+(1-\alpha)x_5,\\ 
E(x_2)=(A\mathbf{x})_2=(1-\alpha - \beta_{1})x_1+0.5x_2+\alpha x_3+ (1-\alpha)x_4+\alpha x_5,\\
E(x_3)=(A\mathbf{x})_3=\alpha x_1 +(1-\alpha)x_2+ 0.5x_3+ (\alpha + \beta_{2})x_4+ (1-\alpha)x_5,\\
E(x_4)=(A\mathbf{x})_4=(1-\alpha)x_1+ \alpha x_2 + (1-\alpha - \beta_{2})x_3 + 0.5 x_4 + \alpha x_5, \\
E(x_5)=(A\mathbf{x})_5=\alpha x_1 + (1-\alpha)x_2 + \alpha x_3 + (1-\alpha)x_4 + 0.5 x_5,\\
\end{aligned}
\end{equation}

and derive a set of four distinct equations $(A\mathbf{x})_1=(A\mathbf{x})_2$, $(A\mathbf{x})_1=(A\mathbf{x})_3$, $(A\mathbf{x})_1=(A\mathbf{x})_4$, and $(A\mathbf{x})_1=(A\mathbf{x})_5$ along with the constraint $x_1+x_2+x_3+x_4+x_5=1$. Hence, we have five equations in five variables $x_1$, $x_2$, $x_3$, $x_4$, and $x_5$. % $\in [0,1]$. 
Solving these five equations, we obtain

%\begin{equation}\label{eq17}
%\begin{aligned}
%x_1 = \frac{(\alpha - 0.5)(\alpha - 0.5 + \beta_{2})}{(\alpha - 0.5)[5(\alpha - 0.5) + 3\beta_{1} + 3\beta_{2}] + \beta_{1}\beta_{2}},\\
%x_2 = \frac{(\alpha - 0.5)(\alpha - 0.5 + \beta_{2})}{(\alpha - 0.5)[5(\alpha - 0.5) + 3\beta_{1} + 3\beta_{2}] + \beta_{1}\beta_{2}},\\
%x_3 = \frac{(\alpha - 0.5)(\alpha - 0.5 + \beta_{1})}{(\alpha - 0.5)[5(\alpha - 0.5) + 3\beta_{1} + 3\beta_{2}] + \beta_{1}\beta_{2}},\\
%x_4 = \frac{(\alpha - 0.5)(\alpha - 0.5 + \beta_{1})}{(\alpha - 0.5)[5(\alpha - 0.5) + 3\beta_{1} + 3\beta_{2}] + \beta_{1}\beta_{2}}, \text{and}\\
%x_5 = \frac{(\alpha - 0.5)(\alpha - 0.5 + \beta_{1} + \beta_{2}) + \beta_{1}\beta_{2}}{(\alpha - 0.5)[5(\alpha - 0.5) + 3\beta_{1} + 3\beta_{2}] + \beta_{1}\beta_{2}}\\
%=\frac{(\alpha - 0.5 + \beta_{2})(\alpha - 0.5 + \beta_{1})}{(\alpha - 0.5)[5(\alpha - 0.5) + 3\beta_{1} + 3\beta_{2}] + \beta_{1}\beta_{2}}.
%%x_{1} = \frac{(\alpha - 0.5 + \beta_{2})}{[5(\alpha - 0.5) + 3\beta_{1} + 3\beta_{2}] + \beta_{1}\beta_{2}},\\
%%x_{2} = \frac{(\alpha - 0.5 + \beta_{2})}{[5(\alpha - 0.5) + 3\beta_{1} + 3\beta_{2}] + \beta_{1}\beta_{2}},\\
%%x_{3}=\frac{(\alpha - 0.5 + \beta_{1})}{[5(\alpha - 0.5) + 3\beta_{1} + 3\beta_{2}] + \beta_{1}\beta_{2}},\\
%%x_{4}=\frac{(\alpha - 0.5 + \beta_{1})}{[5(\alpha - 0.5) + 3\beta_{1} + 3\beta_{2}] + \beta_{1}\beta_{2}}, \text{and}\\
%%x_{5}=\frac{(\alpha - 0.5 + \beta_{1} + \beta_{2}) + \beta_{1}\beta_{2}}{[5(\alpha - 0.5) + 3\beta_{1} + 3\beta_{2}] + \beta_{1}\beta_{2}}.
%\end{aligned}
%\end{equation}

\begin{equation}\label{eq17}
\begin{aligned}
x_1 = x_2 = \frac{d_1(d_1 + \beta_{2})}{d_2},\\
%x_2 = \frac{(\alpha - 0.5)(\alpha - 0.5 + \beta_{2})}{d_2},\\
x_3 = x_4 = \frac{d_1(d_1 + \beta_{1})}{d_2},\\
%x_4 = \frac{(\alpha - 0.5)(\alpha - 0.5 + \beta_{1})}{d_2}, \text{and}\\
x_5 %= \frac{d_1(d_1 + \beta_{1} + \beta_{2}) + \beta_{1}\beta_{2}}{d_2}\\
=\frac{(d_1 + \beta_{2})(d_1 + \beta_{1})}{d_2},
%x_{1} = \frac{(\alpha - 0.5 + \beta_{2})}{[5(\alpha - 0.5) + 3\beta_{1} + 3\beta_{2}] + \beta_{1}\beta_{2}},\\
%x_{2} = \frac{(\alpha - 0.5 + \beta_{2})}{[5(\alpha - 0.5) + 3\beta_{1} + 3\beta_{2}] + \beta_{1}\beta_{2}},\\
%x_{3}=\frac{(\alpha - 0.5 + \beta_{1})}{[5(\alpha - 0.5) + 3\beta_{1} + 3\beta_{2}] + \beta_{1}\beta_{2}},\\
%x_{4}=\frac{(\alpha - 0.5 + \beta_{1})}{[5(\alpha - 0.5) + 3\beta_{1} + 3\beta_{2}] + \beta_{1}\beta_{2}}, \text{and}\\
%x_{5}=\frac{(\alpha - 0.5 + \beta_{1} + \beta_{2}) + \beta_{1}\beta_{2}}{[5(\alpha - 0.5) + 3\beta_{1} + 3\beta_{2}] + \beta_{1}\beta_{2}}.
\end{aligned}
\end{equation}

where $d_1=\alpha - 0.5$, and $d_2=(\alpha - 0.5)[5(\alpha - 0.5) + 3\beta_{1} + 3\beta_{2}] + \beta_{1}\beta_{2}$.

\begin{figure*}[!t]
	\centerline{\includegraphics[width=1.0\textwidth]{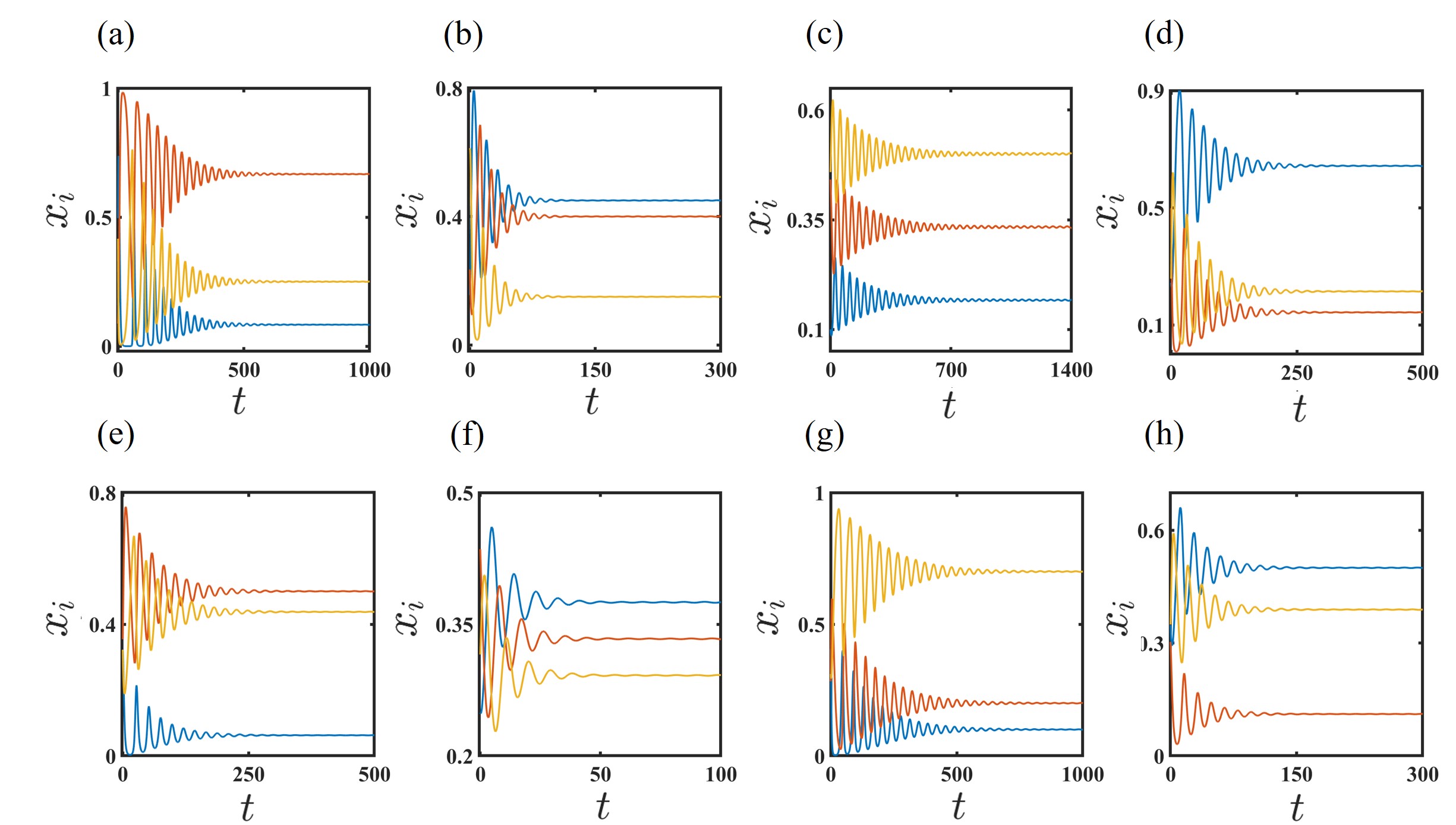}}
	\caption{{\bf Temporal evolution of three species under the triadic interaction for diverse choices of perturbations}: $N=3$ species are interacted as per the interaction matrix $\Tilde{A}$ given in Eq.\ \eqref{eq28} with fixed $\alpha=0.25$. We consider eight different cases, viz.\ (a) $\beta_i > 0$, $i=1,2,3$, (b) $\beta_1,\beta_2>0$, and $\beta_3<0$, (c) $\beta_1>0$, $\beta_2<0$, and $\beta_3>0$, (d) $\beta_1>0$, $\beta_2<0$, and $\beta_3<0$, (e) $\beta_1<0$, $\beta_2>0$, and $\beta_3>0$, (f) $\beta_1<0$, $\beta_2>0$, and $\beta_3<0$, (g) $\beta_1<0$, $\beta_2<0$, and $\beta_3>0$, and (h) $\beta_1<0$, $\beta_2<0$, and $\beta_3<0$. All these possible combinations allow species coexistence with different densities (a) $(0.08333, 0.6667, 0.25)$, (b) $(0.45, 0.4, 0.15)$, (c)  $(0.16667, 0.33333, 0.5)$,  (d) $(0.64286, 0.14286, 0.21429)$, (e)  $(0.0625, 0.5, 0.4375)$, (f)  $ (0.375, 0.33333, 0.29167)$,(g)  $(0.1, 0.2, 0.7)$, and (h)  $(0.5, 0.1111, 0.3889)$, respectively. Here, $|\beta_1|=0.1$, $|\beta_2|=0.15$, and $|\beta_3|=0.2$. The blue, red, and yellow curves contemplate temporal evolutions of the densities $x_1$, $x_2$, and $x_3$, respectively.}
	\label{Picture7}
\end{figure*}

\par This formula is equivalent to the relation \eqref{eq16} for $k=2$ and $\beta_1=\beta_2$. Figure \eqref{Picture5} portrays the impact of two different amounts of perturbations on the first four species among 
$N=5$ 
species with 
$\alpha=0.75$. When 
$\beta_1=\beta_2=0.1$, 
Fig.\ \eqref{Picture5} (a) reflects the decreament of first four species' density due to the employed external perturbations. This outcome attests to the essential signature of external perturbation on the species. The asymptotic values of the 
$x_1=x_2=x_3=x_4=0.18519$, 
and 
$x_5=0.25926$ fits absolutely perfectly with the derived equilibrium \eqref{eq17} of $\Tilde{A}$
(c.f.\ \eqref{eq16}) for 
$\beta_1=\beta_2=0.1$. 
Interestingly, when 
$\beta_2=0.1>0$ 
and 
$\beta_1=-0.1<0$, 
the first two species experience the internal perturbation, and the third and fourth species feel the presence of the external one. The fifth species is left undisturbed. A clear expectation from our previous observation is that
$x_1=x_2$ (orange and blue bar in Fig.\ \eqref{Picture5} (b))
will increase and 
$x_3=x_4$ (purple and brown) 
will decrease compared to the unperturbed case (cyan). The steady state values of the three clusters are $x_1 = 0.28926$, $x_2 = 0.28926$, $x_3 = 0.12397$, $x_4 = 0.12397$, and $x_5 = 0.17355$. Similarly, here we can design the relative density of first (or second) species with the last one as $\Delta_R (1/2-5) = \frac{\alpha - 0.5}{\alpha - 0.5 + \beta_2}$ and the relative density of third (or forth) species with the last one as $\Delta_R (3/4-5) = \frac{\alpha - 0.5}{\alpha - 0.5 + \beta_1}$. Clearly, if we increase the negative strength of $\beta_1 (\beta_2 = 0.1)$, the $\Delta_R (3/4-5)$ will increase, but $\Delta_R (1/2-5)$ will be fixed. Also, if we increase the positive strength of $\beta_2 (\beta_1 = -0.1)$, $\Delta_R (1/2-5)$ will decrease, however $\Delta_R (3/4-5)$ will be fixed.  
%Noticeably, the first two species always
% 
% 
% 
% 
% 
% 
% 
% 
% 
% 
% 
% 
% 
% 
% 
% 
% 
% 
% 
% 
% values  of the three clusters are $x_1=0.28926,x_2=0.28926,x_3=0.12397,x_4=0.12397,x_5=0.17355$.
%Similarly here we can design the relative density of first (or second) species with the last one as $\Delta_R(1/2-5)=\frac{\alpha-0.5+\beta_2}{(\alpha-0.5+\beta_1+\beta_2)+\beta_1\beta_2}$ 
%and the  relative density of third (or forth) species with the last one as
%$\Delta_R(3/4-5)=\frac{\alpha-0.5+\beta_1}{(\alpha-0.5+\beta_1+\beta_2)+\beta_1\beta_2}$.
%Clearly, if we increase the negative strength of 
%$\beta_1$ ($\beta_2=0.1$), 
%the
%$\Delta_R(1/2-5)$ 
%will increase  
%and 
%$\Delta_R(3/4-5)$ 
%will decrease slowly.
%Also, if we increase the positive strength of 
%$\beta_2$ ($\beta_1=-0.1$), 
%both
%$\Delta_R(1/2-5)$ 
%and 
%$\Delta_R(3/4-5)$ 
%will decrease slowly (numerical data not shown here). 
Noticeably, the first two species always win in the competition due to the internal perturbation.

%is demonstrated in Fig.\ \eqref{Picture5} (b).
A similar but reverse scenario is found in Fig.\ \eqref{Picture5} (c), where the external perturbation in the first two species (orange and blue bar)  with $\beta_1=0.1>0$ 
reduces their density effectively. The internal perturbation with $\beta_2=-0.1<0$ 
positively affects the third and fourth species (purple and brown bar), and their densities enhance compared to the undisturbed case (cyan). We find all the species settle down to a three-cluster state with 
$x_3=x_4=0.28926>x_5= 0.17355>x_1=x_2=0.12397$. 
The influence of internal perturbations on the first four species is investigated in Fig.\ \eqref{Picture5} (d) by considering 
$\beta_1=\beta_2=-0.1$. The system \eqref{eq3} converges to the stable stationary point $x_1=x_2=x_3=x_4=0.21739, x_5=0.13043$, which matches excellently with our analytical derivation \eqref{eq17}. The impact of such negative perturbations improves the species density and consequently, diminishes the unperturbed species density 
$x_5$.
%%{\color{black} Here if we increase the negative strength of 
%	$\beta_1$ from 0 to $-0.1$
%	 for $\beta_2=-0.1$, 
%	both
%	$\Delta_R(1/2-5)$ 
%	and 
%	$\Delta_R(3/4-5)$ 
%	will increase, particularly 
%	$\Delta_R(1/2-5)$  will increase rapidly.
%The reverse feature will appear if we negatively increase 
%$\beta_2$ from 0 to $-0.1$ and 
%fix $\beta_2$ at $-0.1$.
%Please note that, we choose 
%$\beta_{1,2}$
%such a way that the network always provides positive feasible equilibrium.
%}

\par Thus, the previous analysis confirms the appearance of three distinct clusters 
$x_1=x_2$, $x_3=x_4$ 
and $x_5$ for non-zero perturbations with 
$\beta_1 \neq \beta_2$. 
To further scrutinize, we plot three two dimensional parameter spaces in Fig.\ \eqref{Picture6} by varying 
$\beta_1$ and $\beta_2$ within 
$[-0.2,0.2]$ for $N=5$. $\alpha$ 
is kept fixed here at 
$0.75$. The obtained results are completely fitted with our derived species densities at Eq. \eqref{eq17}. For a fixed value of $\alpha=0.75$ and fixed value of 
$\beta_1$, $x_1=x_2$ 
will magnify for increament of $\beta_2$ as revealed through Fig.\ \eqref{Picture6} (a). A similar increment of densities in $x_3=x_4$ is observed with enhancement of $\beta_1$ in Fig.\ \eqref{Picture6} (b) for fixed values of $\alpha=0.75$ and $\beta_2$.
Figure \eqref{Picture6} (c) uncovers the reciprocity of 
$\beta_1-\beta_2$ for $\alpha=0.75$ 
for the unperturbed fifth species. 

\par In fact, one can quickly determine the effect of
$k<N$ perturbations 
$\beta_{1}$, $\beta_{2}$, $\cdots$, $\beta_{k}$ 
on first $2k$ out of $N$ species. 
Here, we assume $N$ is odd, and a single species cannot be perturbed more than once. Under these circumstances, the species densities will be derived analytically by calculating the equilibrium of 
$\Tilde{A}$ 
as follows,

\begin{widetext}
\begin{equation}\label{eq18}
\begin{aligned}
x_{1} = x_{2} = \frac{(\alpha - 0.5)(\alpha - 0.5 + \beta_{2})(\alpha - 0.5 + \beta_{3})\cdots(\alpha - 0.5 + \beta_{k})}{\chi},\\
x_{3} = x_{4}= \frac{(\alpha - 0.5)(\alpha - 0.5 + \beta_{1})(\alpha - 0.5 + \beta_{3})\cdots(\alpha - 0.5 + \beta_{k})}{\chi},\\
x_{5} = x_{6} = \frac{(\alpha - 0.5)(\alpha - 0.5 + \beta_{1})(\alpha - 0.5 + \beta_{2})\cdots(\alpha - 0.5 + \beta_{k})}{\chi},\\
\vdots \hspace{5cm}\\
x_{2k-1} = x_{2k} = \frac{(\alpha - 0.5)(\alpha - 0.5 + \beta_{1})(\alpha - 0.5 + \beta_{2})\cdots(\alpha - 0.5 + \beta_{k-1})}{\chi}.
\end{aligned}
\end{equation}
\end{widetext}

Here, 

\begin{equation}\label{eq19}
\begin{aligned}
\chi=(\alpha - 0.5)\cdots [(\alpha - 0.5)[(\alpha - 0.5)[N(\alpha - 0.5) + (N-2) B_{1}] \\+ (N-4) B_{2}] + (N-6) B_{3}] + (N-8) B_{4}] \cdots +(N-2k) B_{k},
\end{aligned}
\end{equation}

where, 

\begin{equation}\label{eq20}
\begin{aligned}
\begin{split}
B_{1} = \sum_{i=1}^{k} \beta_{i},
B_{2} = \sum_{i=1}^{k}\sum_{j\le i} \beta_{i} \beta_{j},\\
B_{3} = \sum_{i=1}^{k} \sum_{j\le i} \sum_{p\le j} \beta_{i}\beta_{j}\beta_{p},
\cdots, \hspace{1.5cm}
B_{k} = \beta_{1}\beta_{2}\beta_{3}\cdots\beta_{k}.
\end{split}
\end{aligned}
\end{equation}

The densities of the remaining $(N-2k)$ unperturbed species are equal and it can be calculated from the constraint $\sum_{i=1}^{N} x_{i} = 1$. Thus, we have

\begin{widetext}
\begin{equation}\label{eq19000}
\begin{aligned}
x_{\rm unperturbed}=\dfrac{(\alpha - 0.5+\beta_1)(\alpha - 0.5+\beta_2)(\alpha - 0.5+\beta_3)\cdots(\alpha - 0.5+\beta_k)}{\chi}.
\end{aligned}
\end{equation}
\end{widetext}

\par More generally relative density of a species having $\beta_l$ perturbation with respect to a unperturbed species can be written as $\Delta_R (\rm Pert(\beta_i)-Unpert) = \frac{\alpha - 0.5}{\alpha - 0.5 + \beta_i}$. So, with the increase of positive strength of $\beta_k$, this relative density will decrease and reverse in the case of an increase in negative strength.

\subsection{Random perturbation in $\dfrac{N(N-1)}{2}$ species-pair (Case: $N=3$)}

\par Still, now we have applied the perturbations so that the same species is not perturbed more than once. What if we disturb the same species more than once? Does it affect the prediction of steady states through the equilibrium of the interaction matrix? To investigate this particular case, we consider $N=3$ species and examine with the following interaction matrix,

\begin{equation}\label{eq28}
\Tilde{A} = 
\begin{bmatrix}
0.5 & \gamma_1 & \gamma_2 \\
1-\gamma_1 & 0.5 & \gamma_3 \\
1-\gamma_2 & 1-\gamma_3 & 0.5
\end{bmatrix}.
\end{equation}

\begin{figure*}[!t]
	\centerline{\includegraphics[width=1.0\textwidth]{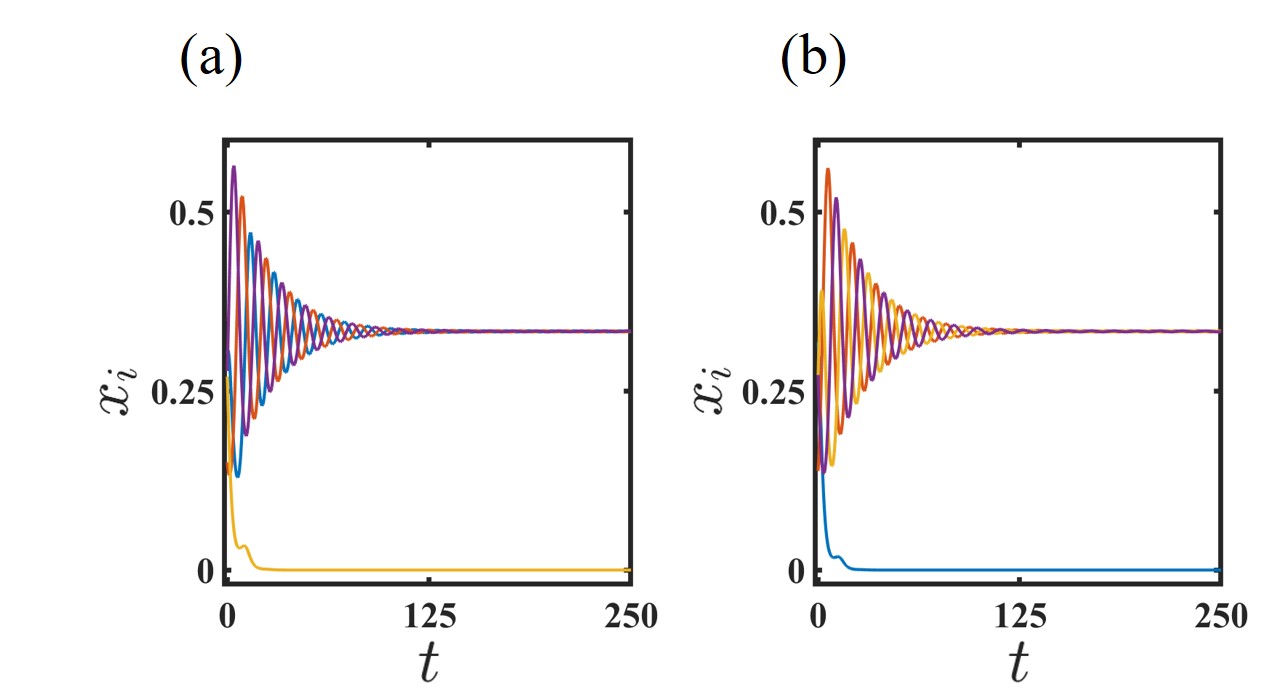}}
	\caption{{\bf Non-existence of an even number of species}: (a) $\alpha=0.75 > 0.5$, and (b) $\alpha=0.25 < 0.5$ lead to the extinction of the fourth and first species, respectively. Both cases allow the coexistence of the remaining $(N-1)$ species with equal densities. Here, $N=4$.}
	\label{Picture8}
\end{figure*}

where $\gamma_1=\alpha+\beta_1$, $\gamma_2=1-\alpha+\beta_2$, and $\gamma_3=\alpha+\beta_3$. Each $\gamma_i$ lies within $[0,1]-\{0.5\}$. We can simulate the equilibrium of this matrix similarly by following the earlier process in Eq.\ \eqref{eq5}, and we obtain

\begin{equation}\label{eq29}
\begin{aligned}
p_{1} = \frac{\nu_3}{\nu}, \hspace{0.2cm}
p_{2} = \frac{\nu_2}{\nu}, \hspace{0.2cm} \text{and} \hspace{0.2cm}
p_3=\frac{\nu_1}{\nu},
\end{aligned}
\end{equation}

where $\nu_1=(\alpha - 0.5)+\beta_{1}$, $\nu_2=(\alpha - 0.5)-\beta_{2}$, $\nu_3=(\alpha - 0.5)+\beta_{3}$, and $\nu=\nu_1+\nu_2+\nu_3=3(\alpha - 0.5) + (\beta_{1} - \beta_{2} + \beta_{3})$.
Here we have fixed $\alpha$ at  $0.25$, and we have considered one particular case where $|\beta_1|=0.1 < |\beta_2|=0.15 < |\beta_3|=0.2$. 
%This derivation perfectly fits the numerically simulated stationary states for all possible combinations of $\beta_1$, $\beta_2$, and $\beta_3$ in Fig.\ \eqref{Picture7}. 
We choose random initial conditions from the permissible range and plot the variation of their corresponding densities (Fig.\ \eqref{Picture7}), including the transient. %We choose $\beta_1 \neq \beta_2 \neq \beta_3$. 
%Since we perturb each individual species more than once in this case, thus we find complicated relationships \eqref{eq29} among the long-term densities of each species depending on the four parameters $\alpha \neq 0.5$, $\beta_i \neq 0$ for $i=1,2,3$. 
Depending upon the condition that $|\beta_1| < |\beta_2| < |\beta_3|$, we 
observe that the order of species density is same for $\beta_1 < 0 $ and $\beta_1 > 0 $ while keeping $\beta_2$ and $\beta_3$ unchanged (see column-wise in Fig.\ \eqref{Picture7} i.e sub-figure "(a) and (e)" or "(b) and (f)" etc.). 
At first, we focus on  the Fig.\ \eqref{Picture7} (a). Here  all $\beta$ s are  positive. Therefore, at $\alpha=0.25$, 
using the Eq.\ \eqref{eq29} we can easily write $|\nu_2|>|\nu_2|>|\nu_3|$. 
%As $\alpha = 0.25$, $(\alpha-0.5)$ is a fixed negative quantity and more over $|(\alpha-0.5)|>|\beta_i|$, for $i=1,2,3$, so the denominator ($\nu$) is also negative. As the denominator is constant comparing the numerator will give the order of species density (the more negative the more density).
%So, when $\beta_i > 0$, for $i=1,2,3$, $|\nu_2|$ will be highest and $\nu_3|$ will be lowest which
This condition leads us to the comparative relation among the species as: $x_1 ({\rm blue}) < x_3 ({\rm yellow}) < x_2 ({\rm red})$. Noticeably, due to the presence of all perturbation, the enhanced cooperation between  the first two species (as $\beta_1<0$) is not helping the first species to gain more density, rather it loses its density (Fig.\ \eqref{Picture7} (a)), however species wins in the competition.
Also in the case of $\beta_1<0$, $\beta_{2,3}>0$,   $|\nu_2| > |\nu_1|$ 
as $|\beta_1| < |\beta_2|$,
and the condition  $|\nu_1| > |\nu_3|$ 
is obvious as the sign of $\beta_1$ is opposite ($<0$) to $\beta_3$ (>0) which makes  $\nu_1$ more negative. 
Thus, the same relation among the species density appears in the Fig.\ \eqref{Picture7}(e).
Now let's consider the case where $\beta_1>0$ (or $\beta_1<0$), $\beta_2<0$, and $\beta_3>0$. Using the \eqref{eq29} we can see that 
$|\nu_1| > |\nu_2|>|\nu_3|$ 
and hence we observe the species density having the order $x_1 < x_2 < x_3$. 
Using the same logic along with \eqref{eq29} we can explain the scenarios for another particular order of $\beta_1$, $\beta_2$ and $\beta_3$.

\section{Conclusion: Current status and future prospects}

\par{\color{black} To understand the maintenance of biodiversity observed in a realistic scenario, recently, an increasing trend has been observed to study ecological systems in the light of higher-order interaction (HOI) \cite{bairey2016high,grilli2017higher, mayfield2017higher, levine2017beyond, singh2021higher,majhi2022dynamics}. Letten et al.\ \cite{letten2019mechanistic} have not only shown how these models approximate the ecological system with better accuracy but also provided motivation for how this indirect effect arises from the phenomenological models. So this paves the way for choosing HOI in our proposed model. It is still debated whether pairwise interactions are distinct from HOI and how to define and implement these interactions. Levine et al.\ \cite{levine2017beyond} have argued that both interactions operate over different timescales, making them distinct, but sometimes both timescales are comparable \cite{grilli2017higher}. Actually, looking at this through the eyes of a mechanical model of resource competition, no distinction can be found \cite{letten2019mechanistic}. Hence, we have considered only the HOI to describe the model.
	
\par Furthermore, in the present article, HOIs are fully characterized by the pairwise interaction without adding new parameters and making this model empirically testable \cite{grilli2017higher}. On the other hand, in ecological systems, RPS dynamics are very well studied. It is essential from the perspective of describing intransitive cycles, cyclic dominance of species, and also describing the theory of species co-existence and diversity\cite{buss1979competitive,sinervo1996rock,paquin1983relative,kerr2002local,allesina2011competitive}. Hence, we developed our model based on generalized RPS game which is really in the center of interest for ecologists. The present manuscript reflects how a network of interacting species can lead to species co-existence and diversity. Also, how changing one interaction can lead to different indirect effects in other species is characterized here.} 
In the presence of higher-order interaction, we investigate the structurally perturbed intransitive cycle of $N$ species. We have determined analytically how the globally stable coexisting equilibrium points are arranged for a long-term evaluation. We have found that when two species reduce their competition (winning probability is decreased slightly from the stronger species), their density increases compared to other species. The analytically calculated globally stable stationary points match perfectly with the numerical derivations. We reveal how the feasible perturbation in the interaction matrix splits the long-time behavior of the replicator equations into a few synchronized clusters. {\color{black} Finally, these results indicate the interaction of how weighted network structures between species can determine the population density in the ecological dynamics \cite{sodhan2021metapopulation}. Internal perturbation is essentially the positive evolution of weaker species, so the winning probability of stronger decreases. Similarly, external perturbation is the positive evolution of stronger, so the winning likelihood of stronger increases. We have shown how the interplay of this evolution leads to diverse species density in a system. Biologists can further test this for competing bacteria in laboratory conditions \cite{friedman2017community}.}

%%%%%%%%%%%%%%%%%%%%%%%%%%%%%%%%%%%%%%%%%%%%%%%%%%%%

\par Future work relies on relaxation of the assumption like species interacting with a particular strength and constraint on the perturbation range and number of perturbations can be given.{\color{black} This will allow us to arrive at the results in a more general setting. Now, depending on the environmental changes, this perturbation can exceed the range we have considered in this model. This will lead to a reversal in dominance, ultimately implying the extinction of some species, which will be more realistic in the large ecological system.} One may wish to look at the spread of the stationary points in a large system when the perturbation is chosen from a truncated normal distribution with having a fixed interaction matrix and finally in a more general case where the interaction matrix is also a random matrix \cite{may1972will,holling1973resilience,allesina2012stability,grilli2017higher}.
{\color{black}This spread of equilibrium points corresponds to different species having very diverse species densities. One may further consider those fundamental concepts like resource consumption and other vast range of physical parameters that can help better understand how small perturbations, i.e., evolution or environmental changes, can drive a species to extension. We conclude with the hope that our perceived theoretical model offers a feasible step toward revealing the influence of higher-order interactions on ecological communities, and our mathematical convenience may be pertinent to much applied biological realism. }

\section*{Appendix} 

\subsection*{Dynamics of even species} \label{Even}

\par Nevertheless, when $N \geq 4$ is even, then it is impossible to calculate the equilibrium of ${A}$ given in Eq.\ \eqref{eq4}. For example, let us consider $N=4$. Then the interaction matrix is given by

\begin{equation}\label{eq21}
{A} = 
\begin{bmatrix}
0.5 & \alpha & 1-\alpha & \alpha \\
1-\alpha & 0.5 & \alpha & 1-\alpha \\
\alpha & 1-\alpha & 0.5 & \alpha \\
1-\alpha & \alpha & 1-\alpha & 0.5
\end{bmatrix}.
\end{equation}

We omit the case of $\alpha=0.5$ to avoid multiple optimal strategies. Proceeding as the previous way given in Eq.\ \eqref{eq5}, we get

\begin{equation}\label{eq22}
\begin{aligned}
E(x_1)=(0.5-\alpha)p_1+(1-2\alpha)p_3+\alpha,\\
E(x_2)=(\alpha-0.5)p_2+(2\alpha-1)p_3+(1-\alpha),\\
E(x_3)=(1-2\alpha)p_2+(0.5-\alpha)p_3+\alpha,\\
E(x_4)=(0.5-\alpha)p_1+(\alpha-0.5)p_2+(0.5-\alpha)p_3+0.5,
\end{aligned}
\end{equation}

where we use $p_1+p_2+p_3+p_4=1$ to eliminate $p_4$. To evaluate the equilibrium of the above matrix $A$, we equate each of these expected values %. Hence, we have 

\begin{equation}\label{eq23}
E(x_1)=E(x_2)=E(x_3)=E(x_4).
\end{equation}

For $E(x_1)=E(x_4)$, we get 

\begin{equation}\label{eq24}
p_2+p_3=1,
\end{equation}

since $\alpha \neq 0.5$. Similarly, for $E(x_2)=E(x_3)$, we get 

\begin{equation}\label{eq25}
p_2+p_3=\frac{2}{3}.
\end{equation}

\begin{figure*}[!t]
	\centerline{\includegraphics[width=1.0\textwidth]{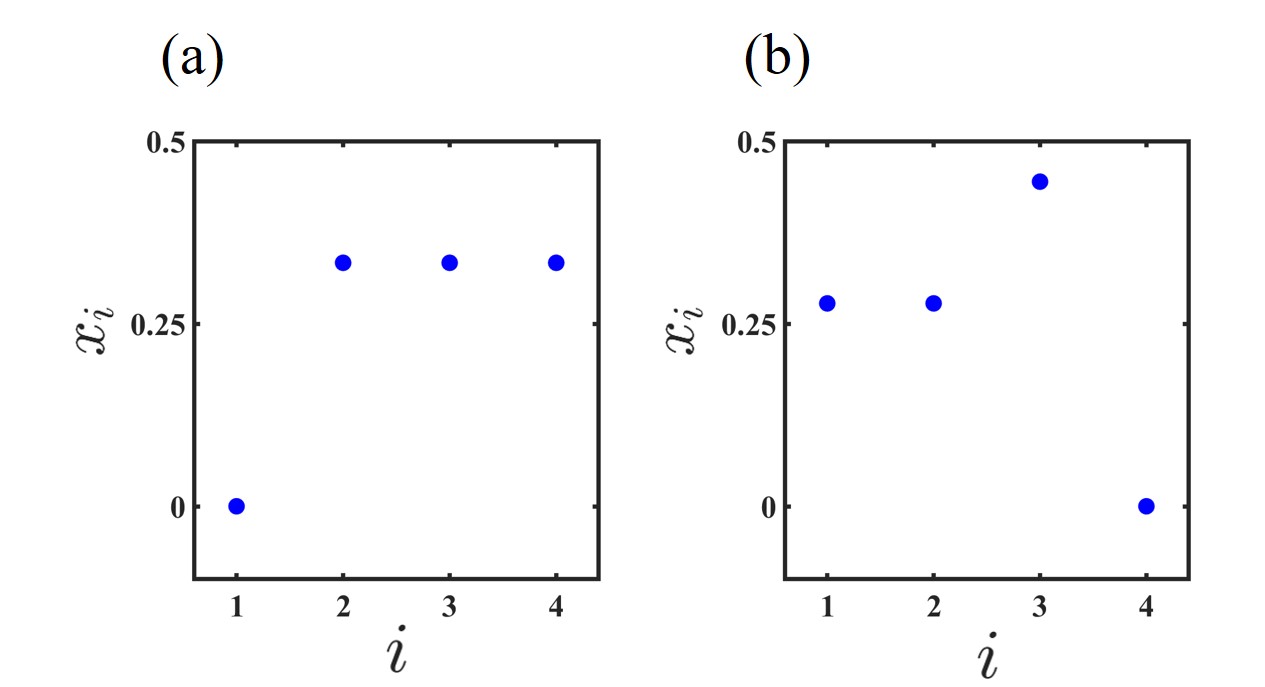}}
	\caption{{\bf Effect of localized perturbation on even number species}: We choose $N=4$ species and apply the perturbation of strength $\beta=0.15$ on the interaction of first-second species. Depending on the choice $\alpha \neq 0.5$, the system converges to different species densities. (a) When $\alpha=0.25<0.5$, the first species goes extinct similar to the unperturbed case ($\beta=0$), as shown in Fig.\ \eqref{Picture8} (b). The system settles down to the stationary point $\big(0,\frac{1}{3},\frac{1}{3},\frac{1}{3}\big)$. (b) The $\alpha=0.75>0.5$ leads to the extinction of the fourth species in the long run. The remaining three species will survive maintaining Eq.\ \eqref{eq12}, and the system \eqref{eq3} converges to $(0.2778,0.2778,0.4444,0)$. The result is independent of the choice of initial conditions chosen from the permissible domain. }
	\label{Picture9}
\end{figure*}

This clearly indicates that Eq.\ \eqref{eq23} is a set of inconsistent equations. Hence, few of these $p_i$'s must be zero. But, as soon as any one of these $p_i$ is zero, we have a system of $(N-1)$ species, and since $N$ is even, $(N-1)$ becomes odd. Thus, as per our previous analysis, the system settles down to a stable equilibrium with one zero component and $(N-1)$ non-zero component. Our numerical simulation suggests depending on the choice of $\alpha \neq 0.5$, either the first or the $N$-th species will die when $N$ is even. For $\alpha > 0.5$, we have 

\begin{equation}\label{eq26}
\begin{aligned}
x_i=\dfrac{1}{N-1}, \text{for} \hspace{0.2cm} i = 1,2,3,\cdots,N-1\\
x_N=0. \hspace{2cm}
\end{aligned}
\end{equation}

Similarly, we obtain for $\alpha < 0.5$ the following stationary point,

\begin{equation}\label{eq27}
\begin{aligned}
x_1=0, \hspace{2cm} \\
x_i=\dfrac{1}{N-1}, \text{for} \hspace{0.2cm} i=2,3,\cdots,N
\end{aligned}
\end{equation}

Figure \eqref{Picture8} demonstrates the extinction of one species and the remaining $(N-1)=3$ species coexist with equal densities $\frac{1}{N-1}=\frac{1}{3}$ irrespective of the choice of $\alpha \neq 0.5$. Figure \eqref{Picture8}(a) %In the subfigure (a) 
for $\alpha = 0.75$, the fourth species dies out in the long term, and in Fig.\ \eqref{Picture8} (b) for $\alpha = 0.25$, the first species extincts. The choice of initial conditions maintaining the constraint $x_1(0)+x_2(0)+x_3(0)=1$ does not affect the final outcome. Thus, when $N$ is even, any one of the species (either first or the $N$-th depending on the $\alpha$) die out, and the prediction of the stationary point containing the remaining $(N-1)$ species can be quickly done using the equilibrium of the interaction matrix $A$, as $(N-1)$ is odd.

\par We consider the following approach to prove the impossibility of gaining a feasible solution with $A$ of even order. Let us consider the matrix $C$ with entries 

%\begin{equation}
%C=circ\big(0, 1, -1, \cdots).
%\label{eq30}
%\end{equation}

%where,

\begin{equation}
C_{ij}=\begin{cases}
0 \quad &\text{if} \, i=j \\
1 \quad &\text{if} \, A_{ij}>0.5 \\
-1 \quad &\text{if} \, A_{ij}<0.5 \\\end{cases}
\label{eq31}
\end{equation}

The entries of the matrix $C$ can be interpreted in the way that if $i$-th species win over the $j$-th, then we assign $C_{ij}=1$. That is in the case of a directed graph; if the arc is pointed from $i$ to $j$, we have $C_{ij}=1$. Similarly, if the arc is from $j$ to $i$, i.e., $i$ is dominated by $j$, then we assign $C_{ji}=-1$. The diagonal entries are considered as zero since there is no own competition. Let $\mathbf{p}=\big(p_1,p_2,p_3,\cdots,p_N \big)^T$ be the strategy vector with $\sum_{i=1}^{N}p_i=1$ and $p_i \geq 0$ $\forall$ $i$. To calculate the optimal strategy, i.e., the non-zero solution $\mathbf{p}$, we must have

\begin{equation}
C\mathbf{p}=\mathbf{0}\\
\implies detC=0.
\label{eq32}
\end{equation}

Since $C$ is a skew symmetric matrix, if $N$ is even, then $detC=1$ is a non-zero perfect square. Hence, there does not exist any non-zero solution $\mathbf{p}$  of \eqref{eq32}. On the other hand, the determinant of a skew symmetric matrix of odd order equals zero. So, calculation of the non-zero $\mathbf{p}$ with $\sum_{i=1}^{N}p_i=1$ and $p_i \geq 0$ $\forall$ $i$ is possible, only if $N \geq 3$ is odd.

\subsection*{Perturbation in single species for even number of species}

\par We now consider the impact of perturbation on an even number of species. For simplicity, we choose $N=4$. When $\alpha=0.25<0.5$, the influence of $\beta>0$ is neutralized as the first species die out. The remaining species coexist with equal densities as reflected through Fig.\ \eqref{Picture9} (a). However, $\alpha>0.5$ and $\beta>0$ reduces $x_1$ and $x_2$ similar to the effect of external perturbations on triadic interaction as shown in Fig.\ \eqref{Picture3} (b). We choose $\beta=0.15$ and $\alpha=0.75$ in Fig.\ \eqref{Picture9} (b). The tripple-wise interaction in this case leads to the cessation of fourth species' density.  $x_4$ converges to zero in the long run and positive $\beta$ acts like an external perturbation on the remaining three species for $\alpha>0.5$. The asymptotic values of $x_1,x_2,$ and $x_3$ can be predicted using the theoretically calculated equilibrium in Eq.\ \eqref{eq12} due to the extinction of the fourth species. The survivability of the first three species now becomes equivalent to studying the system \eqref{eq3} with $(N-1)=3$ species.

\section*{AUTHORS’ CONTRIBUTIONS}

Both the authors Sourin Chatterjee and Sayantan Nag Chowdhury have contributed equally to this work.\\
{\bf Sourin Chatterjee:} Conceptualization, Methodology,
Software, Validation, Formal analysis, Investigation Visualization, Writing - review \& editing. {\bf Sayantan Nag Chowdhury:}  Methodology, Validation, Formal analysis, Investigation,  Writing, Visualization. {\bf Dibakar Ghosh:}  Validation, Visualization, Writing - review \& editing. {\bf Chittaranjan Hens:} Conceptualization, Methodology, Software,  Validation, Formal analysis, Investigation, Writing, Visualization.

\section*{Competing interests.} We declare we have no competing interests.

\section*{ACKNOWLEDGMENTS}
    
{\color{black}The authors gratefully acknowledge the anonymous referees for their insightful suggestions that helped considerably improve the manuscript. We are indebted to Jacopo Grilli of The Abdus Salam International Centre for Theoretical Physics for valuable discussions.} Sayantan Nag Chowdhury would like to acknowledge the CSIR (Project No. 09/093(0194)/2020-EMR-I) for financial assistance. {\color{black}Sayantan Nag Chowdhury also wants to thank the Department of Science and Technology, Govt. of India, for the financial support through Grant No. NMICPS/006/MD/2020-21 during the end of this work.} Chittaranjan Hens is financially supported by the INSPIRE-Faculty grant (Code: IFA17-PH193).

\section*{DATA AVAILABILITY}
The data supporting this study's findings are openly available on Github at \cite{web_1}.

\bibliographystyle{apsrev4-1}
\bibliography{HOI_Bib}

\begin{thebibliography}{69}%
\makeatletter
\providecommand \@ifxundefined [1]{%
 \@ifx{#1\undefined}
}%
\providecommand \@ifnum [1]{%
 \ifnum #1\expandafter \@firstoftwo
 \else \expandafter \@secondoftwo
 \fi
}%
\providecommand \@ifx [1]{%
 \ifx #1\expandafter \@firstoftwo
 \else \expandafter \@secondoftwo
 \fi
}%
\providecommand \natexlab [1]{#1}%
\providecommand \enquote  [1]{``#1''}%
\providecommand \bibnamefont  [1]{#1}%
\providecommand \bibfnamefont [1]{#1}%
\providecommand \citenamefont [1]{#1}%
\providecommand \href@noop [0]{\@secondoftwo}%
\providecommand \href [0]{\begingroup \@sanitize@url \@href}%
\providecommand \@href[1]{\@@startlink{#1}\@@href}%
\providecommand \@@href[1]{\endgroup#1\@@endlink}%
\providecommand \@sanitize@url [0]{\catcode `\\12\catcode `\$12\catcode
  `\&12\catcode `\#12\catcode `\^12\catcode `\_12\catcode `\%12\relax}%
\providecommand \@@startlink[1]{}%
\providecommand \@@endlink[0]{}%
\providecommand \url  [0]{\begingroup\@sanitize@url \@url }%
\providecommand \@url [1]{\endgroup\@href {#1}{\urlprefix }}%
\providecommand \urlprefix  [0]{URL }%
\providecommand \Eprint [0]{\href }%
\providecommand \doibase [0]{http://dx.doi.org/}%
\providecommand \selectlanguage [0]{\@gobble}%
\providecommand \bibinfo  [0]{\@secondoftwo}%
\providecommand \bibfield  [0]{\@secondoftwo}%
\providecommand \translation [1]{[#1]}%
\providecommand \BibitemOpen [0]{}%
\providecommand \bibitemStop [0]{}%
\providecommand \bibitemNoStop [0]{.\EOS\space}%
\providecommand \EOS [0]{\spacefactor3000\relax}%
\providecommand \BibitemShut  [1]{\csname bibitem#1\endcsname}%
\let\auto@bib@innerbib\@empty
%</preamble>
\bibitem [{\citenamefont {Grilli}\ \emph {et~al.}(2017)\citenamefont {Grilli},
  \citenamefont {Barab{\'a}s}, \citenamefont {Michalska-Smith},\ and\
  \citenamefont {Allesina}}]{grilli2017higher}%
  \BibitemOpen
  \bibfield  {author} {\bibinfo {author} {\bibfnamefont {J.}~\bibnamefont
  {Grilli}}, \bibinfo {author} {\bibfnamefont {G.}~\bibnamefont {Barab{\'a}s}},
  \bibinfo {author} {\bibfnamefont {M.~J.}\ \bibnamefont {Michalska-Smith}}, \
  and\ \bibinfo {author} {\bibfnamefont {S.}~\bibnamefont {Allesina}},\ }\href
  {\doibase https://doi.org/10.1038/nature23273} {\bibfield  {journal}
  {\bibinfo  {journal} {Nature}\ }\textbf {\bibinfo {volume} {548}},\ \bibinfo
  {pages} {210} (\bibinfo {year} {2017})}\BibitemShut {NoStop}%
\bibitem [{\citenamefont {Lotka}(1920)}]{lotka1920analytical}%
  \BibitemOpen
  \bibfield  {author} {\bibinfo {author} {\bibfnamefont {A.~J.}\ \bibnamefont
  {Lotka}},\ }\href {\doibase https://doi.org/10.1073/pnas.6.7.410} {\bibfield
  {journal} {\bibinfo  {journal} {Proceedings of the National Academy of
  Sciences}\ }\textbf {\bibinfo {volume} {6}},\ \bibinfo {pages} {410}
  (\bibinfo {year} {1920})}\BibitemShut {NoStop}%
\bibitem [{\citenamefont {Volterra}(1926)}]{volterra1926variations}%
  \BibitemOpen
  \bibfield  {author} {\bibinfo {author} {\bibfnamefont {V.}~\bibnamefont
  {Volterra}},\ }\href {\doibase https://doi.org/10.1093/icesjms/3.1.3}
  {\bibfield  {journal} {\bibinfo  {journal} {Animal Ecology}\ ,\ \bibinfo
  {pages} {409}} (\bibinfo {year} {1926})}\BibitemShut {NoStop}%
\bibitem [{\citenamefont {May}\ and\ \citenamefont
  {Leonard}(1975)}]{may1975nonlinear}%
  \BibitemOpen
  \bibfield  {author} {\bibinfo {author} {\bibfnamefont {R.~M.}\ \bibnamefont
  {May}}\ and\ \bibinfo {author} {\bibfnamefont {W.~J.}\ \bibnamefont
  {Leonard}},\ }\href {\doibase http://dx.doi.org/10.1137/0129022} {\bibfield
  {journal} {\bibinfo  {journal} {SIAM Journal on Applied Mathematics}\
  }\textbf {\bibinfo {volume} {29}},\ \bibinfo {pages} {243} (\bibinfo {year}
  {1975})}\BibitemShut {NoStop}%
\bibitem [{\citenamefont {Nag~Chowdhury}\ \emph
  {et~al.}(2021{\natexlab{a}})\citenamefont {Nag~Chowdhury}, \citenamefont
  {Kundu}, \citenamefont {Perc},\ and\ \citenamefont
  {Ghosh}}]{chowdhury2021complex}%
  \BibitemOpen
  \bibfield  {author} {\bibinfo {author} {\bibfnamefont {S.}~\bibnamefont
  {Nag~Chowdhury}}, \bibinfo {author} {\bibfnamefont {S.}~\bibnamefont
  {Kundu}}, \bibinfo {author} {\bibfnamefont {M.}~\bibnamefont {Perc}}, \ and\
  \bibinfo {author} {\bibfnamefont {D.}~\bibnamefont {Ghosh}},\ }\href
  {\doibase https://doi.org/10.1098/rspa.2021.0397} {\bibfield  {journal}
  {\bibinfo  {journal} {Proceedings of the Royal Society A}\ }\textbf {\bibinfo
  {volume} {477}},\ \bibinfo {pages} {20210397} (\bibinfo {year}
  {2021}{\natexlab{a}})}\BibitemShut {NoStop}%
\bibitem [{\citenamefont {Hubbell}(2011)}]{hubbell2011unified}%
  \BibitemOpen
  \bibfield  {author} {\bibinfo {author} {\bibfnamefont {S.~P.}\ \bibnamefont
  {Hubbell}},\ }\href {\doibase https://doi.org/10.1515/9781400837526} {\emph
  {\bibinfo {title} {The unified neutral theory of biodiversity and
  biogeography (MPB-32)}}}\ (\bibinfo  {publisher} {Princeton University
  Press},\ \bibinfo {year} {2011})\BibitemShut {NoStop}%
\bibitem [{\citenamefont {May}(1974)}]{may1974biological}%
  \BibitemOpen
  \bibfield  {author} {\bibinfo {author} {\bibfnamefont {R.~M.}\ \bibnamefont
  {May}},\ }\href {\doibase https://doi.org/10.1126/science.186.4164.645}
  {\bibfield  {journal} {\bibinfo  {journal} {Science}\ }\textbf {\bibinfo
  {volume} {186}},\ \bibinfo {pages} {645} (\bibinfo {year}
  {1974})}\BibitemShut {NoStop}%
\bibitem [{\citenamefont {Nag~Chowdhury}\ \emph
  {et~al.}(2021{\natexlab{b}})\citenamefont {Nag~Chowdhury}, \citenamefont
  {Kundu}, \citenamefont {Banerjee}, \citenamefont {Perc},\ and\ \citenamefont
  {Ghosh}}]{chowdhury2021eco}%
  \BibitemOpen
  \bibfield  {author} {\bibinfo {author} {\bibfnamefont {S.}~\bibnamefont
  {Nag~Chowdhury}}, \bibinfo {author} {\bibfnamefont {S.}~\bibnamefont
  {Kundu}}, \bibinfo {author} {\bibfnamefont {J.}~\bibnamefont {Banerjee}},
  \bibinfo {author} {\bibfnamefont {M.}~\bibnamefont {Perc}}, \ and\ \bibinfo
  {author} {\bibfnamefont {D.}~\bibnamefont {Ghosh}},\ }\href {\doibase
  https://doi.org/10.1016/j.jtbi.2021.110606} {\bibfield  {journal} {\bibinfo
  {journal} {Journal of Theoretical Biology}\ }\textbf {\bibinfo {volume}
  {518}},\ \bibinfo {pages} {110606} (\bibinfo {year}
  {2021}{\natexlab{b}})}\BibitemShut {NoStop}%
\bibitem [{\citenamefont {May}(2019)}]{may2019stability}%
  \BibitemOpen
  \bibfield  {author} {\bibinfo {author} {\bibfnamefont {R.~M.}\ \bibnamefont
  {May}},\ }\href {\doibase https://doi.org/10.1515/9780691206912} {\emph
  {\bibinfo {title} {Stability and complexity in model ecosystems}}}\ (\bibinfo
   {publisher} {Princeton university press},\ \bibinfo {year}
  {2019})\BibitemShut {NoStop}%
\bibitem [{\citenamefont {Chesson}(2000)}]{chesson2000mechanisms}%
  \BibitemOpen
  \bibfield  {author} {\bibinfo {author} {\bibfnamefont {P.}~\bibnamefont
  {Chesson}},\ }\href {\doibase
  https://doi.org/10.1146/annurev.ecolsys.31.1.343} {\bibfield  {journal}
  {\bibinfo  {journal} {Annual review of Ecology and Systematics}\ }\textbf
  {\bibinfo {volume} {31}},\ \bibinfo {pages} {343} (\bibinfo {year}
  {2000})}\BibitemShut {NoStop}%
\bibitem [{\citenamefont {Nag~Chowdhury}\ \emph {et~al.}(2020)\citenamefont
  {Nag~Chowdhury}, \citenamefont {Kundu}, \citenamefont {Duh}, \citenamefont
  {Perc},\ and\ \citenamefont {Ghosh}}]{nag2020cooperation}%
  \BibitemOpen
  \bibfield  {author} {\bibinfo {author} {\bibfnamefont {S.}~\bibnamefont
  {Nag~Chowdhury}}, \bibinfo {author} {\bibfnamefont {S.}~\bibnamefont
  {Kundu}}, \bibinfo {author} {\bibfnamefont {M.}~\bibnamefont {Duh}}, \bibinfo
  {author} {\bibfnamefont {M.}~\bibnamefont {Perc}}, \ and\ \bibinfo {author}
  {\bibfnamefont {D.}~\bibnamefont {Ghosh}},\ }\href {\doibase
  https://doi.org/10.3390/e22040485} {\bibfield  {journal} {\bibinfo  {journal}
  {Entropy}\ }\textbf {\bibinfo {volume} {22}},\ \bibinfo {pages} {485}
  (\bibinfo {year} {2020})}\BibitemShut {NoStop}%
\bibitem [{\citenamefont {Gao}\ \emph {et~al.}(2016)\citenamefont {Gao},
  \citenamefont {Barzel},\ and\ \citenamefont
  {Barab{\'a}si}}]{gao2016universal}%
  \BibitemOpen
  \bibfield  {author} {\bibinfo {author} {\bibfnamefont {J.}~\bibnamefont
  {Gao}}, \bibinfo {author} {\bibfnamefont {B.}~\bibnamefont {Barzel}}, \ and\
  \bibinfo {author} {\bibfnamefont {A.-L.}\ \bibnamefont {Barab{\'a}si}},\
  }\href {\doibase https://doi.org/10.1038/nature16948} {\bibfield  {journal}
  {\bibinfo  {journal} {Nature}\ }\textbf {\bibinfo {volume} {530}},\ \bibinfo
  {pages} {307} (\bibinfo {year} {2016})}\BibitemShut {NoStop}%
\bibitem [{\citenamefont {Hens}\ \emph {et~al.}(2019)\citenamefont {Hens},
  \citenamefont {Harush}, \citenamefont {Haber}, \citenamefont {Cohen},\ and\
  \citenamefont {Barzel}}]{hens2019spatiotemporal}%
  \BibitemOpen
  \bibfield  {author} {\bibinfo {author} {\bibfnamefont {C.}~\bibnamefont
  {Hens}}, \bibinfo {author} {\bibfnamefont {U.}~\bibnamefont {Harush}},
  \bibinfo {author} {\bibfnamefont {S.}~\bibnamefont {Haber}}, \bibinfo
  {author} {\bibfnamefont {R.}~\bibnamefont {Cohen}}, \ and\ \bibinfo {author}
  {\bibfnamefont {B.}~\bibnamefont {Barzel}},\ }\href {\doibase
  https://doi.org/10.1038/s41567-018-0409-0} {\bibfield  {journal} {\bibinfo
  {journal} {Nature Physics}\ }\textbf {\bibinfo {volume} {15}},\ \bibinfo
  {pages} {403} (\bibinfo {year} {2019})}\BibitemShut {NoStop}%
\bibitem [{\citenamefont {Roy}\ \emph {et~al.}(2022)\citenamefont {Roy},
  \citenamefont {Nag~Chowdhury}, \citenamefont {Mali}, \citenamefont {Perc},\
  and\ \citenamefont {Ghosh}}]{roy2022multigames}%
  \BibitemOpen
  \bibfield  {author} {\bibinfo {author} {\bibfnamefont {S.}~\bibnamefont
  {Roy}}, \bibinfo {author} {\bibfnamefont {S.}~\bibnamefont {Nag~Chowdhury}},
  \bibinfo {author} {\bibfnamefont {P.~C.}\ \bibnamefont {Mali}}, \bibinfo
  {author} {\bibfnamefont {M.}~\bibnamefont {Perc}}, \ and\ \bibinfo {author}
  {\bibfnamefont {D.}~\bibnamefont {Ghosh}},\ }\href {\doibase
  https://doi.org/10.1371/journal.pone.0272719} {\bibfield  {journal} {\bibinfo
   {journal} {PLOS ONE}\ }\textbf {\bibinfo {volume} {17}},\ \bibinfo {pages}
  {1} (\bibinfo {year} {2022})}\BibitemShut {NoStop}%
\bibitem [{\citenamefont {Chase}\ and\ \citenamefont
  {Leibold}(2009)}]{chase2009ecological}%
  \BibitemOpen
  \bibfield  {author} {\bibinfo {author} {\bibfnamefont {J.~M.}\ \bibnamefont
  {Chase}}\ and\ \bibinfo {author} {\bibfnamefont {M.~A.}\ \bibnamefont
  {Leibold}},\ }\href {\doibase
  https://doi.org/10.7208/chicago/9780226101811.001.0001} {\emph {\bibinfo
  {title} {Ecological Niches}}}\ (\bibinfo  {publisher} {University of Chicago
  Press},\ \bibinfo {year} {2009})\BibitemShut {NoStop}%
\bibitem [{\citenamefont {Sober{\'o}n}\ and\ \citenamefont
  {Nakamura}(2009)}]{soberon2009niches}%
  \BibitemOpen
  \bibfield  {author} {\bibinfo {author} {\bibfnamefont {J.}~\bibnamefont
  {Sober{\'o}n}}\ and\ \bibinfo {author} {\bibfnamefont {M.}~\bibnamefont
  {Nakamura}},\ }\href {\doibase http://dx.doi.org/10.1073/pnas.0901637106}
  {\bibfield  {journal} {\bibinfo  {journal} {Proceedings of the National
  Academy of Sciences}\ }\textbf {\bibinfo {volume} {106}},\ \bibinfo {pages}
  {19644} (\bibinfo {year} {2009})}\BibitemShut {NoStop}%
\bibitem [{\citenamefont {Tilman}(2004)}]{tilman2004niche}%
  \BibitemOpen
  \bibfield  {author} {\bibinfo {author} {\bibfnamefont {D.}~\bibnamefont
  {Tilman}},\ }\href {\doibase https://doi.org/10.1073/pnas.0403458101}
  {\bibfield  {journal} {\bibinfo  {journal} {Proceedings of the National
  Academy of Sciences}\ }\textbf {\bibinfo {volume} {101}},\ \bibinfo {pages}
  {10854} (\bibinfo {year} {2004})}\BibitemShut {NoStop}%
\bibitem [{\citenamefont {Bell}(2000)}]{bell2000distribution}%
  \BibitemOpen
  \bibfield  {author} {\bibinfo {author} {\bibfnamefont {G.}~\bibnamefont
  {Bell}},\ }\href {\doibase https://doi.org/10.1086/303345} {\bibfield
  {journal} {\bibinfo  {journal} {The American Naturalist}\ }\textbf {\bibinfo
  {volume} {155}},\ \bibinfo {pages} {606} (\bibinfo {year}
  {2000})}\BibitemShut {NoStop}%
\bibitem [{\citenamefont {Brokaw}\ and\ \citenamefont
  {Busing}(2000)}]{brokaw2000niche}%
  \BibitemOpen
  \bibfield  {author} {\bibinfo {author} {\bibfnamefont {N.}~\bibnamefont
  {Brokaw}}\ and\ \bibinfo {author} {\bibfnamefont {R.~T.}\ \bibnamefont
  {Busing}},\ }\href {\doibase https://doi.org/10.1016/s0169-5347(00)01822-x}
  {\bibfield  {journal} {\bibinfo  {journal} {Trends in Ecology \& Evolution}\
  }\textbf {\bibinfo {volume} {15}},\ \bibinfo {pages} {183} (\bibinfo {year}
  {2000})}\BibitemShut {NoStop}%
\bibitem [{\citenamefont {Wennekes}\ \emph {et~al.}(2012)\citenamefont
  {Wennekes}, \citenamefont {Rosindell},\ and\ \citenamefont
  {Etienne}}]{wennekes2012neutral}%
  \BibitemOpen
  \bibfield  {author} {\bibinfo {author} {\bibfnamefont {P.~L.}\ \bibnamefont
  {Wennekes}}, \bibinfo {author} {\bibfnamefont {J.}~\bibnamefont {Rosindell}},
  \ and\ \bibinfo {author} {\bibfnamefont {R.~S.}\ \bibnamefont {Etienne}},\
  }\href {\doibase http://dx.doi.org/10.1007/s10441-012-9144-6} {\bibfield
  {journal} {\bibinfo  {journal} {Acta Biotheoretica}\ }\textbf {\bibinfo
  {volume} {60}},\ \bibinfo {pages} {257} (\bibinfo {year} {2012})}\BibitemShut
  {NoStop}%
\bibitem [{\citenamefont {Chisholm}\ and\ \citenamefont
  {Pacala}(2010)}]{chisholm2010niche}%
  \BibitemOpen
  \bibfield  {author} {\bibinfo {author} {\bibfnamefont {R.~A.}\ \bibnamefont
  {Chisholm}}\ and\ \bibinfo {author} {\bibfnamefont {S.~W.}\ \bibnamefont
  {Pacala}},\ }\href {\doibase https://doi.org/10.1073/pnas.1009387107}
  {\bibfield  {journal} {\bibinfo  {journal} {Proceedings of the National
  Academy of Sciences}\ }\textbf {\bibinfo {volume} {107}},\ \bibinfo {pages}
  {15821} (\bibinfo {year} {2010})}\BibitemShut {NoStop}%
\bibitem [{\citenamefont {Gravel}\ \emph {et~al.}(2006)\citenamefont {Gravel},
  \citenamefont {Canham}, \citenamefont {Beaudet},\ and\ \citenamefont
  {Messier}}]{gravel2006reconciling}%
  \BibitemOpen
  \bibfield  {author} {\bibinfo {author} {\bibfnamefont {D.}~\bibnamefont
  {Gravel}}, \bibinfo {author} {\bibfnamefont {C.~D.}\ \bibnamefont {Canham}},
  \bibinfo {author} {\bibfnamefont {M.}~\bibnamefont {Beaudet}}, \ and\
  \bibinfo {author} {\bibfnamefont {C.}~\bibnamefont {Messier}},\ }\href
  {\doibase https://doi.org/10.1111/j.1461-0248.2006.00884.x} {\bibfield
  {journal} {\bibinfo  {journal} {Ecology Letters}\ }\textbf {\bibinfo {volume}
  {9}},\ \bibinfo {pages} {399} (\bibinfo {year} {2006})}\BibitemShut {NoStop}%
\bibitem [{\citenamefont {McGill}(2003)}]{mcgill2003test}%
  \BibitemOpen
  \bibfield  {author} {\bibinfo {author} {\bibfnamefont {B.~J.}\ \bibnamefont
  {McGill}},\ }\href {\doibase https://doi.org/10.1038/nature01583} {\bibfield
  {journal} {\bibinfo  {journal} {Nature}\ }\textbf {\bibinfo {volume} {422}},\
  \bibinfo {pages} {881} (\bibinfo {year} {2003})}\BibitemShut {NoStop}%
\bibitem [{\citenamefont {Clark}\ and\ \citenamefont
  {McLachlan}(2003)}]{clark2003stability}%
  \BibitemOpen
  \bibfield  {author} {\bibinfo {author} {\bibfnamefont {J.~S.}\ \bibnamefont
  {Clark}}\ and\ \bibinfo {author} {\bibfnamefont {J.~S.}\ \bibnamefont
  {McLachlan}},\ }\href {\doibase https://doi.org/10.1038/nature01632}
  {\bibfield  {journal} {\bibinfo  {journal} {Nature}\ }\textbf {\bibinfo
  {volume} {423}},\ \bibinfo {pages} {635} (\bibinfo {year}
  {2003})}\BibitemShut {NoStop}%
\bibitem [{\citenamefont {Huisman}\ and\ \citenamefont
  {Weissing}(1999)}]{huisman1999biodiversity}%
  \BibitemOpen
  \bibfield  {author} {\bibinfo {author} {\bibfnamefont {J.}~\bibnamefont
  {Huisman}}\ and\ \bibinfo {author} {\bibfnamefont {F.~J.}\ \bibnamefont
  {Weissing}},\ }\href {\doibase https://doi.org/10.1038/46540} {\bibfield
  {journal} {\bibinfo  {journal} {Nature}\ }\textbf {\bibinfo {volume} {402}},\
  \bibinfo {pages} {407} (\bibinfo {year} {1999})}\BibitemShut {NoStop}%
\bibitem [{\citenamefont {He}\ \emph {et~al.}(2011)\citenamefont {He},
  \citenamefont {Mobilia},\ and\ \citenamefont
  {T{\"a}uber}}]{he2011coexistence}%
  \BibitemOpen
  \bibfield  {author} {\bibinfo {author} {\bibfnamefont {Q.}~\bibnamefont
  {He}}, \bibinfo {author} {\bibfnamefont {M.}~\bibnamefont {Mobilia}}, \ and\
  \bibinfo {author} {\bibfnamefont {U.~C.}\ \bibnamefont {T{\"a}uber}},\ }\href
  {\doibase https://doi.org/10.1140/epjb/e2011-20259-x} {\bibfield  {journal}
  {\bibinfo  {journal} {The European Physical Journal B}\ }\textbf {\bibinfo
  {volume} {82}},\ \bibinfo {pages} {97} (\bibinfo {year} {2011})}\BibitemShut
  {NoStop}%
\bibitem [{\citenamefont {Szab{\'o}}\ and\ \citenamefont
  {Fath}(2007)}]{szabo2007evolutionary}%
  \BibitemOpen
  \bibfield  {author} {\bibinfo {author} {\bibfnamefont {G.}~\bibnamefont
  {Szab{\'o}}}\ and\ \bibinfo {author} {\bibfnamefont {G.}~\bibnamefont
  {Fath}},\ }\href {\doibase https://doi.org/10.1016/j.physrep.2007.04.004}
  {\bibfield  {journal} {\bibinfo  {journal} {Physics Reports}\ }\textbf
  {\bibinfo {volume} {446}},\ \bibinfo {pages} {97} (\bibinfo {year}
  {2007})}\BibitemShut {NoStop}%
\bibitem [{\citenamefont {Berr}\ \emph {et~al.}(2009)\citenamefont {Berr},
  \citenamefont {Reichenbach}, \citenamefont {Schottenloher},\ and\
  \citenamefont {Frey}}]{berr2009zero}%
  \BibitemOpen
  \bibfield  {author} {\bibinfo {author} {\bibfnamefont {M.}~\bibnamefont
  {Berr}}, \bibinfo {author} {\bibfnamefont {T.}~\bibnamefont {Reichenbach}},
  \bibinfo {author} {\bibfnamefont {M.}~\bibnamefont {Schottenloher}}, \ and\
  \bibinfo {author} {\bibfnamefont {E.}~\bibnamefont {Frey}},\ }\href {\doibase
  https://doi.org/10.1103/PhysRevLett.102.048102} {\bibfield  {journal}
  {\bibinfo  {journal} {Physical Review Letters}\ }\textbf {\bibinfo {volume}
  {102}},\ \bibinfo {pages} {048102} (\bibinfo {year} {2009})}\BibitemShut
  {NoStop}%
\bibitem [{\citenamefont {Sinervo}\ and\ \citenamefont
  {Lively}(1996)}]{sinervo1996rock}%
  \BibitemOpen
  \bibfield  {author} {\bibinfo {author} {\bibfnamefont {B.}~\bibnamefont
  {Sinervo}}\ and\ \bibinfo {author} {\bibfnamefont {C.~M.}\ \bibnamefont
  {Lively}},\ }\href {\doibase https://doi.org/10.1038/380240a0} {\bibfield
  {journal} {\bibinfo  {journal} {Nature}\ }\textbf {\bibinfo {volume} {380}},\
  \bibinfo {pages} {240} (\bibinfo {year} {1996})}\BibitemShut {NoStop}%
\bibitem [{\citenamefont {Kerr}\ \emph {et~al.}(2002)\citenamefont {Kerr},
  \citenamefont {Riley}, \citenamefont {Feldman},\ and\ \citenamefont
  {Bohannan}}]{kerr2002local}%
  \BibitemOpen
  \bibfield  {author} {\bibinfo {author} {\bibfnamefont {B.}~\bibnamefont
  {Kerr}}, \bibinfo {author} {\bibfnamefont {M.~A.}\ \bibnamefont {Riley}},
  \bibinfo {author} {\bibfnamefont {M.~W.}\ \bibnamefont {Feldman}}, \ and\
  \bibinfo {author} {\bibfnamefont {B.~J.}\ \bibnamefont {Bohannan}},\ }\href
  {\doibase https://doi.org/10.1038/nature00823} {\bibfield  {journal}
  {\bibinfo  {journal} {Nature}\ }\textbf {\bibinfo {volume} {418}},\ \bibinfo
  {pages} {171} (\bibinfo {year} {2002})}\BibitemShut {NoStop}%
\bibitem [{\citenamefont {Laird}\ and\ \citenamefont
  {Schamp}(2006)}]{laird2006competitive}%
  \BibitemOpen
  \bibfield  {author} {\bibinfo {author} {\bibfnamefont {R.~A.}\ \bibnamefont
  {Laird}}\ and\ \bibinfo {author} {\bibfnamefont {B.~S.}\ \bibnamefont
  {Schamp}},\ }\href {\doibase https://doi.org/10.1086/506259} {\bibfield
  {journal} {\bibinfo  {journal} {The American Naturalist}\ }\textbf {\bibinfo
  {volume} {168}},\ \bibinfo {pages} {182} (\bibinfo {year}
  {2006})}\BibitemShut {NoStop}%
\bibitem [{\citenamefont {Hibbing}\ \emph {et~al.}(2010)\citenamefont
  {Hibbing}, \citenamefont {Fuqua}, \citenamefont {Parsek},\ and\ \citenamefont
  {Peterson}}]{hibbing2010bacterial}%
  \BibitemOpen
  \bibfield  {author} {\bibinfo {author} {\bibfnamefont {M.~E.}\ \bibnamefont
  {Hibbing}}, \bibinfo {author} {\bibfnamefont {C.}~\bibnamefont {Fuqua}},
  \bibinfo {author} {\bibfnamefont {M.~R.}\ \bibnamefont {Parsek}}, \ and\
  \bibinfo {author} {\bibfnamefont {S.~B.}\ \bibnamefont {Peterson}},\ }\href
  {\doibase https://doi.org/10.1038/nrmicro2259} {\bibfield  {journal}
  {\bibinfo  {journal} {Nature Reviews Microbiology}\ }\textbf {\bibinfo
  {volume} {8}},\ \bibinfo {pages} {15} (\bibinfo {year} {2010})}\BibitemShut
  {NoStop}%
\bibitem [{\citenamefont {Gilpin}(1975)}]{gilpin1975limit}%
  \BibitemOpen
  \bibfield  {author} {\bibinfo {author} {\bibfnamefont {M.~E.}\ \bibnamefont
  {Gilpin}},\ }\href {\doibase https://doi.org/10.1086/282973} {\bibfield
  {journal} {\bibinfo  {journal} {The American Naturalist}\ }\textbf {\bibinfo
  {volume} {109}},\ \bibinfo {pages} {51} (\bibinfo {year} {1975})}\BibitemShut
  {NoStop}%
\bibitem [{\citenamefont {Allesina}\ and\ \citenamefont
  {Levine}(2011)}]{allesina2011competitive}%
  \BibitemOpen
  \bibfield  {author} {\bibinfo {author} {\bibfnamefont {S.}~\bibnamefont
  {Allesina}}\ and\ \bibinfo {author} {\bibfnamefont {J.~M.}\ \bibnamefont
  {Levine}},\ }\href {\doibase https://doi.org/10.1073/pnas.1014428108}
  {\bibfield  {journal} {\bibinfo  {journal} {Proceedings of the National
  Academy of Sciences}\ }\textbf {\bibinfo {volume} {108}},\ \bibinfo {pages}
  {5638} (\bibinfo {year} {2011})}\BibitemShut {NoStop}%
\bibitem [{\citenamefont {Bhattacharyya}\ \emph {et~al.}(2020)\citenamefont
  {Bhattacharyya}, \citenamefont {Sinha}, \citenamefont {De},\ and\
  \citenamefont {Hens}}]{bhattacharyya2020mortality}%
  \BibitemOpen
  \bibfield  {author} {\bibinfo {author} {\bibfnamefont {S.}~\bibnamefont
  {Bhattacharyya}}, \bibinfo {author} {\bibfnamefont {P.}~\bibnamefont
  {Sinha}}, \bibinfo {author} {\bibfnamefont {R.}~\bibnamefont {De}}, \ and\
  \bibinfo {author} {\bibfnamefont {C.}~\bibnamefont {Hens}},\ }\href {\doibase
  https://doi.org/10.1103/PhysRevE.102.012220} {\bibfield  {journal} {\bibinfo
  {journal} {Physical Review E}\ }\textbf {\bibinfo {volume} {102}},\ \bibinfo
  {pages} {012220} (\bibinfo {year} {2020})}\BibitemShut {NoStop}%
\bibitem [{\citenamefont {Szolnoki}\ \emph {et~al.}(2014)\citenamefont
  {Szolnoki}, \citenamefont {Mobilia}, \citenamefont {Jiang}, \citenamefont
  {Szczesny}, \citenamefont {Rucklidge},\ and\ \citenamefont
  {Perc}}]{szolnoki2014cyclic}%
  \BibitemOpen
  \bibfield  {author} {\bibinfo {author} {\bibfnamefont {A.}~\bibnamefont
  {Szolnoki}}, \bibinfo {author} {\bibfnamefont {M.}~\bibnamefont {Mobilia}},
  \bibinfo {author} {\bibfnamefont {L.-L.}\ \bibnamefont {Jiang}}, \bibinfo
  {author} {\bibfnamefont {B.}~\bibnamefont {Szczesny}}, \bibinfo {author}
  {\bibfnamefont {A.~M.}\ \bibnamefont {Rucklidge}}, \ and\ \bibinfo {author}
  {\bibfnamefont {M.}~\bibnamefont {Perc}},\ }\href {\doibase
  https://doi.org/10.1098/rsif.2014.0735} {\bibfield  {journal} {\bibinfo
  {journal} {Journal of the Royal Society Interface}\ }\textbf {\bibinfo
  {volume} {11}},\ \bibinfo {pages} {20140735} (\bibinfo {year}
  {2014})}\BibitemShut {NoStop}%
\bibitem [{\citenamefont {Reichenbach}\ \emph {et~al.}(2007)\citenamefont
  {Reichenbach}, \citenamefont {Mobilia},\ and\ \citenamefont
  {Frey}}]{reichenbach2007mobility}%
  \BibitemOpen
  \bibfield  {author} {\bibinfo {author} {\bibfnamefont {T.}~\bibnamefont
  {Reichenbach}}, \bibinfo {author} {\bibfnamefont {M.}~\bibnamefont
  {Mobilia}}, \ and\ \bibinfo {author} {\bibfnamefont {E.}~\bibnamefont
  {Frey}},\ }\href {\doibase http://dx.doi.org/10.1038/nature06095} {\bibfield
  {journal} {\bibinfo  {journal} {Nature}\ }\textbf {\bibinfo {volume} {448}},\
  \bibinfo {pages} {1046} (\bibinfo {year} {2007})}\BibitemShut {NoStop}%
\bibitem [{\citenamefont {Islam}\ \emph {et~al.}(2022)\citenamefont {Islam},
  \citenamefont {Mondal}, \citenamefont {Mobilia}, \citenamefont
  {Bhattacharyya},\ and\ \citenamefont {Hens}}]{islam2022effect}%
  \BibitemOpen
  \bibfield  {author} {\bibinfo {author} {\bibfnamefont {S.}~\bibnamefont
  {Islam}}, \bibinfo {author} {\bibfnamefont {A.}~\bibnamefont {Mondal}},
  \bibinfo {author} {\bibfnamefont {M.}~\bibnamefont {Mobilia}}, \bibinfo
  {author} {\bibfnamefont {S.}~\bibnamefont {Bhattacharyya}}, \ and\ \bibinfo
  {author} {\bibfnamefont {C.}~\bibnamefont {Hens}},\ }\href {\doibase
  https://doi.org/10.1103/PhysRevE.105.014215} {\bibfield  {journal} {\bibinfo
  {journal} {Physical Review E}\ }\textbf {\bibinfo {volume} {105}},\ \bibinfo
  {pages} {014215} (\bibinfo {year} {2022})}\BibitemShut {NoStop}%
\bibitem [{\citenamefont {Park}\ \emph {et~al.}(2017)\citenamefont {Park},
  \citenamefont {Do}, \citenamefont {Jang},\ and\ \citenamefont
  {Lai}}]{park2017emergence}%
  \BibitemOpen
  \bibfield  {author} {\bibinfo {author} {\bibfnamefont {J.}~\bibnamefont
  {Park}}, \bibinfo {author} {\bibfnamefont {Y.}~\bibnamefont {Do}}, \bibinfo
  {author} {\bibfnamefont {B.}~\bibnamefont {Jang}}, \ and\ \bibinfo {author}
  {\bibfnamefont {Y.-C.}\ \bibnamefont {Lai}},\ }\href {\doibase
  https://doi.org/10.1038/s41598-017-07911-4} {\bibfield  {journal} {\bibinfo
  {journal} {Scientific Reports}\ }\textbf {\bibinfo {volume} {7}},\ \bibinfo
  {pages} {1} (\bibinfo {year} {2017})}\BibitemShut {NoStop}%
\bibitem [{\citenamefont {Shi}\ \emph {et~al.}(2010)\citenamefont {Shi},
  \citenamefont {Wang}, \citenamefont {Yang},\ and\ \citenamefont
  {Lai}}]{shi2010basins}%
  \BibitemOpen
  \bibfield  {author} {\bibinfo {author} {\bibfnamefont {H.}~\bibnamefont
  {Shi}}, \bibinfo {author} {\bibfnamefont {W.-X.}\ \bibnamefont {Wang}},
  \bibinfo {author} {\bibfnamefont {R.}~\bibnamefont {Yang}}, \ and\ \bibinfo
  {author} {\bibfnamefont {Y.-C.}\ \bibnamefont {Lai}},\ }\href {\doibase
  http://dx.doi.org/10.1103/PhysRevE.81.030901} {\bibfield  {journal} {\bibinfo
   {journal} {Physical Review E}\ }\textbf {\bibinfo {volume} {81}},\ \bibinfo
  {pages} {030901} (\bibinfo {year} {2010})}\BibitemShut {NoStop}%
\bibitem [{\citenamefont {Kirkup}\ and\ \citenamefont
  {Riley}(2004)}]{kirkup2004antibiotic}%
  \BibitemOpen
  \bibfield  {author} {\bibinfo {author} {\bibfnamefont {B.~C.}\ \bibnamefont
  {Kirkup}}\ and\ \bibinfo {author} {\bibfnamefont {M.~A.}\ \bibnamefont
  {Riley}},\ }\href {\doibase https://doi.org/10.1038/nature02429} {\bibfield
  {journal} {\bibinfo  {journal} {Nature}\ }\textbf {\bibinfo {volume} {428}},\
  \bibinfo {pages} {412} (\bibinfo {year} {2004})}\BibitemShut {NoStop}%
\bibitem [{\citenamefont {Soliveres}\ \emph {et~al.}(2015)\citenamefont
  {Soliveres}, \citenamefont {Maestre}, \citenamefont {Ulrich}, \citenamefont
  {Manning}, \citenamefont {Boch}, \citenamefont {Bowker}, \citenamefont
  {Prati}, \citenamefont {Delgado-Baquerizo}, \citenamefont {Quero},
  \citenamefont {Sch{\"o}ning} \emph {et~al.}}]{soliveres2015intransitive}%
  \BibitemOpen
  \bibfield  {author} {\bibinfo {author} {\bibfnamefont {S.}~\bibnamefont
  {Soliveres}}, \bibinfo {author} {\bibfnamefont {F.~T.}\ \bibnamefont
  {Maestre}}, \bibinfo {author} {\bibfnamefont {W.}~\bibnamefont {Ulrich}},
  \bibinfo {author} {\bibfnamefont {P.}~\bibnamefont {Manning}}, \bibinfo
  {author} {\bibfnamefont {S.}~\bibnamefont {Boch}}, \bibinfo {author}
  {\bibfnamefont {M.~A.}\ \bibnamefont {Bowker}}, \bibinfo {author}
  {\bibfnamefont {D.}~\bibnamefont {Prati}}, \bibinfo {author} {\bibfnamefont
  {M.}~\bibnamefont {Delgado-Baquerizo}}, \bibinfo {author} {\bibfnamefont
  {J.~L.}\ \bibnamefont {Quero}}, \bibinfo {author} {\bibfnamefont
  {I.}~\bibnamefont {Sch{\"o}ning}},  \emph {et~al.},\ }\href {\doibase
  https://doi.org/10.1111/ele.12456} {\bibfield  {journal} {\bibinfo  {journal}
  {Ecology Letters}\ }\textbf {\bibinfo {volume} {18}},\ \bibinfo {pages} {790}
  (\bibinfo {year} {2015})}\BibitemShut {NoStop}%
\bibitem [{\citenamefont {Cameron}\ \emph {et~al.}(2009)\citenamefont
  {Cameron}, \citenamefont {White},\ and\ \citenamefont
  {Antonovics}}]{cameron2009parasite}%
  \BibitemOpen
  \bibfield  {author} {\bibinfo {author} {\bibfnamefont {D.~D.}\ \bibnamefont
  {Cameron}}, \bibinfo {author} {\bibfnamefont {A.}~\bibnamefont {White}}, \
  and\ \bibinfo {author} {\bibfnamefont {J.}~\bibnamefont {Antonovics}},\
  }\href {\doibase http://dx.doi.org/10.1111/j.1365-2745.2009.01568.x}
  {\bibfield  {journal} {\bibinfo  {journal} {Journal of Ecology}\ }\textbf
  {\bibinfo {volume} {97}},\ \bibinfo {pages} {1311} (\bibinfo {year}
  {2009})}\BibitemShut {NoStop}%
\bibitem [{\citenamefont {Battiston}\ \emph {et~al.}(2020)\citenamefont
  {Battiston}, \citenamefont {Cencetti}, \citenamefont {Iacopini},
  \citenamefont {Latora}, \citenamefont {Lucas}, \citenamefont {Patania},
  \citenamefont {Young},\ and\ \citenamefont {Petri}}]{battiston2020networks}%
  \BibitemOpen
  \bibfield  {author} {\bibinfo {author} {\bibfnamefont {F.}~\bibnamefont
  {Battiston}}, \bibinfo {author} {\bibfnamefont {G.}~\bibnamefont {Cencetti}},
  \bibinfo {author} {\bibfnamefont {I.}~\bibnamefont {Iacopini}}, \bibinfo
  {author} {\bibfnamefont {V.}~\bibnamefont {Latora}}, \bibinfo {author}
  {\bibfnamefont {M.}~\bibnamefont {Lucas}}, \bibinfo {author} {\bibfnamefont
  {A.}~\bibnamefont {Patania}}, \bibinfo {author} {\bibfnamefont {J.-G.}\
  \bibnamefont {Young}}, \ and\ \bibinfo {author} {\bibfnamefont
  {G.}~\bibnamefont {Petri}},\ }\href {\doibase
  https://doi.org/10.1016/j.physrep.2020.05.004} {\bibfield  {journal}
  {\bibinfo  {journal} {Physics Reports}\ }\textbf {\bibinfo {volume} {874}},\
  \bibinfo {pages} {1} (\bibinfo {year} {2020})}\BibitemShut {NoStop}%
\bibitem [{\citenamefont {Letten}\ and\ \citenamefont
  {Stouffer}(2019)}]{letten2019mechanistic}%
  \BibitemOpen
  \bibfield  {author} {\bibinfo {author} {\bibfnamefont {A.~D.}\ \bibnamefont
  {Letten}}\ and\ \bibinfo {author} {\bibfnamefont {D.~B.}\ \bibnamefont
  {Stouffer}},\ }\href {\doibase https://doi.org/10.1111/ele.13211} {\bibfield
  {journal} {\bibinfo  {journal} {Ecology Letters}\ }\textbf {\bibinfo {volume}
  {22}},\ \bibinfo {pages} {423} (\bibinfo {year} {2019})}\BibitemShut
  {NoStop}%
\bibitem [{\citenamefont {Mayfield}\ and\ \citenamefont
  {Stouffer}(2017)}]{mayfield2017higher}%
  \BibitemOpen
  \bibfield  {author} {\bibinfo {author} {\bibfnamefont {M.~M.}\ \bibnamefont
  {Mayfield}}\ and\ \bibinfo {author} {\bibfnamefont {D.~B.}\ \bibnamefont
  {Stouffer}},\ }\href {\doibase https://doi.org/10.1038/s41559-016-0062}
  {\bibfield  {journal} {\bibinfo  {journal} {Nature ecology \& evolution}\
  }\textbf {\bibinfo {volume} {1}},\ \bibinfo {pages} {1} (\bibinfo {year}
  {2017})}\BibitemShut {NoStop}%
\bibitem [{\citenamefont {Levine}\ \emph {et~al.}(2017)\citenamefont {Levine},
  \citenamefont {Bascompte}, \citenamefont {Adler},\ and\ \citenamefont
  {Allesina}}]{levine2017beyond}%
  \BibitemOpen
  \bibfield  {author} {\bibinfo {author} {\bibfnamefont {J.~M.}\ \bibnamefont
  {Levine}}, \bibinfo {author} {\bibfnamefont {J.}~\bibnamefont {Bascompte}},
  \bibinfo {author} {\bibfnamefont {P.~B.}\ \bibnamefont {Adler}}, \ and\
  \bibinfo {author} {\bibfnamefont {S.}~\bibnamefont {Allesina}},\ }\href
  {\doibase https://doi.org/10.1038/nature22898} {\bibfield  {journal}
  {\bibinfo  {journal} {Nature}\ }\textbf {\bibinfo {volume} {546}},\ \bibinfo
  {pages} {56} (\bibinfo {year} {2017})}\BibitemShut {NoStop}%
\bibitem [{\citenamefont {Bairey}\ \emph {et~al.}(2016)\citenamefont {Bairey},
  \citenamefont {Kelsic},\ and\ \citenamefont {Kishony}}]{bairey2016high}%
  \BibitemOpen
  \bibfield  {author} {\bibinfo {author} {\bibfnamefont {E.}~\bibnamefont
  {Bairey}}, \bibinfo {author} {\bibfnamefont {E.~D.}\ \bibnamefont {Kelsic}},
  \ and\ \bibinfo {author} {\bibfnamefont {R.}~\bibnamefont {Kishony}},\ }\href
  {\doibase https://doi.org/10.1038/ncomms12285} {\bibfield  {journal}
  {\bibinfo  {journal} {Nature Communications}\ }\textbf {\bibinfo {volume}
  {7}},\ \bibinfo {pages} {1} (\bibinfo {year} {2016})}\BibitemShut {NoStop}%
\bibitem [{\citenamefont {Abrams}(1983)}]{abrams1983arguments}%
  \BibitemOpen
  \bibfield  {author} {\bibinfo {author} {\bibfnamefont {P.~A.}\ \bibnamefont
  {Abrams}},\ }\href {\doibase https://doi.org/10.1086/284111} {\bibfield
  {journal} {\bibinfo  {journal} {The American Naturalist}\ }\textbf {\bibinfo
  {volume} {121}},\ \bibinfo {pages} {887} (\bibinfo {year}
  {1983})}\BibitemShut {NoStop}%
\bibitem [{\citenamefont {Lambiotte}\ \emph {et~al.}(2019)\citenamefont
  {Lambiotte}, \citenamefont {Rosvall},\ and\ \citenamefont
  {Scholtes}}]{lambiotte2019networks}%
  \BibitemOpen
  \bibfield  {author} {\bibinfo {author} {\bibfnamefont {R.}~\bibnamefont
  {Lambiotte}}, \bibinfo {author} {\bibfnamefont {M.}~\bibnamefont {Rosvall}},
  \ and\ \bibinfo {author} {\bibfnamefont {I.}~\bibnamefont {Scholtes}},\
  }\href {\doibase https://doi.org/10.1038/s41567-019-0459-y} {\bibfield
  {journal} {\bibinfo  {journal} {Nature physics}\ }\textbf {\bibinfo {volume}
  {15}},\ \bibinfo {pages} {313} (\bibinfo {year} {2019})}\BibitemShut
  {NoStop}%
\bibitem [{\citenamefont {Stouffer}\ \emph {et~al.}(2018)\citenamefont
  {Stouffer}, \citenamefont {Wainwright}, \citenamefont {Flanagan},\ and\
  \citenamefont {Mayfield}}]{stouffer2018cyclic}%
  \BibitemOpen
  \bibfield  {author} {\bibinfo {author} {\bibfnamefont {D.~B.}\ \bibnamefont
  {Stouffer}}, \bibinfo {author} {\bibfnamefont {C.~E.}\ \bibnamefont
  {Wainwright}}, \bibinfo {author} {\bibfnamefont {T.}~\bibnamefont
  {Flanagan}}, \ and\ \bibinfo {author} {\bibfnamefont {M.~M.}\ \bibnamefont
  {Mayfield}},\ }\href {\doibase https://doi.org/10.1111/1365-2745.12960}
  {\bibfield  {journal} {\bibinfo  {journal} {Journal of Ecology}\ }\textbf
  {\bibinfo {volume} {106}},\ \bibinfo {pages} {838} (\bibinfo {year}
  {2018})}\BibitemShut {NoStop}%
\bibitem [{\citenamefont {Li}\ \emph {et~al.}(2020)\citenamefont {Li},
  \citenamefont {Bearup},\ and\ \citenamefont {Liao}}]{li2020habitat}%
  \BibitemOpen
  \bibfield  {author} {\bibinfo {author} {\bibfnamefont {Y.}~\bibnamefont
  {Li}}, \bibinfo {author} {\bibfnamefont {D.}~\bibnamefont {Bearup}}, \ and\
  \bibinfo {author} {\bibfnamefont {J.}~\bibnamefont {Liao}},\ }\href {\doibase
  https://doi.org/10.1098/rspb.2020.1571} {\bibfield  {journal} {\bibinfo
  {journal} {Proceedings of the Royal Society B}\ }\textbf {\bibinfo {volume}
  {287}},\ \bibinfo {pages} {20201571} (\bibinfo {year} {2020})}\BibitemShut
  {NoStop}%
\bibitem [{\citenamefont {Singh}\ and\ \citenamefont
  {Baruah}(2021)}]{singh2021higher}%
  \BibitemOpen
  \bibfield  {author} {\bibinfo {author} {\bibfnamefont {P.}~\bibnamefont
  {Singh}}\ and\ \bibinfo {author} {\bibfnamefont {G.}~\bibnamefont {Baruah}},\
  }\href {\doibase https://doi.org/10.1016/j.physrep.2020.05.004} {\bibfield
  {journal} {\bibinfo  {journal} {Theoretical Ecology}\ }\textbf {\bibinfo
  {volume} {14}},\ \bibinfo {pages} {71} (\bibinfo {year} {2021})}\BibitemShut
  {NoStop}%
\bibitem [{\citenamefont {Abrams}(2000)}]{abrams2000evolution}%
  \BibitemOpen
  \bibfield  {author} {\bibinfo {author} {\bibfnamefont {P.~A.}\ \bibnamefont
  {Abrams}},\ }\href {\doibase https://doi.org/10.1146/annurev.ecolsys.31.1.79}
  {\bibfield  {journal} {\bibinfo  {journal} {Annual Review of Ecology and
  Systematics}\ }\textbf {\bibinfo {volume} {31}},\ \bibinfo {pages} {79}
  (\bibinfo {year} {2000})}\BibitemShut {NoStop}%
\bibitem [{\citenamefont {Spaniel}(2014)}]{spaniel2014game}%
  \BibitemOpen
  \bibfield  {author} {\bibinfo {author} {\bibfnamefont {W.}~\bibnamefont
  {Spaniel}},\ }\href@noop {} {\emph {\bibinfo {title} {Game theory 101: the
  complete textbook}}}\ (\bibinfo  {publisher} {CreateSpace},\ \bibinfo {year}
  {2014})\BibitemShut {NoStop}%
\bibitem [{\citenamefont {Fisher}\ and\ \citenamefont
  {Reeves}(1995)}]{fisher1995optimal}%
  \BibitemOpen
  \bibfield  {author} {\bibinfo {author} {\bibfnamefont {D.~C.}\ \bibnamefont
  {Fisher}}\ and\ \bibinfo {author} {\bibfnamefont {R.~B.}\ \bibnamefont
  {Reeves}},\ }\href {\doibase https://doi.org/10.1016/0024-3795(94)00212-V}
  {\bibfield  {journal} {\bibinfo  {journal} {Linear Algebra and its
  Applications}\ }\textbf {\bibinfo {volume} {217}},\ \bibinfo {pages} {83}
  (\bibinfo {year} {1995})}\BibitemShut {NoStop}%
\bibitem [{\citenamefont {Jensen}(1906)}]{jensen1906fonctions}%
  \BibitemOpen
  \bibfield  {author} {\bibinfo {author} {\bibfnamefont {J.~L. W.~V.}\
  \bibnamefont {Jensen}},\ }\href {\doibase https://doi.org/10.1007/BF02418571}
  {\bibfield  {journal} {\bibinfo  {journal} {Acta Mathematica}\ }\textbf
  {\bibinfo {volume} {30}},\ \bibinfo {pages} {175} (\bibinfo {year}
  {1906})}\BibitemShut {NoStop}%
\bibitem [{\citenamefont {Dekking}\ \emph {et~al.}(2005)\citenamefont
  {Dekking}, \citenamefont {Kraaikamp}, \citenamefont {Lopuha{\"a}},\ and\
  \citenamefont {Meester}}]{dekking2005modern}%
  \BibitemOpen
  \bibfield  {author} {\bibinfo {author} {\bibfnamefont {F.~M.}\ \bibnamefont
  {Dekking}}, \bibinfo {author} {\bibfnamefont {C.}~\bibnamefont {Kraaikamp}},
  \bibinfo {author} {\bibfnamefont {H.~P.}\ \bibnamefont {Lopuha{\"a}}}, \ and\
  \bibinfo {author} {\bibfnamefont {L.~E.}\ \bibnamefont {Meester}},\ }\href
  {\doibase http://dx.doi.org/10.1007/1-84628-168-7} {\emph {\bibinfo {title}
  {A Modern Introduction to Probability and Statistics: Understanding why and
  how}}},\ Vol.\ \bibinfo {volume} {488}\ (\bibinfo  {publisher} {Springer},\
  \bibinfo {year} {2005})\BibitemShut {NoStop}%
\bibitem [{\citenamefont {Ruel}\ and\ \citenamefont
  {Ayres}(1999)}]{ruel1999jensen}%
  \BibitemOpen
  \bibfield  {author} {\bibinfo {author} {\bibfnamefont {J.~J.}\ \bibnamefont
  {Ruel}}\ and\ \bibinfo {author} {\bibfnamefont {M.~P.}\ \bibnamefont
  {Ayres}},\ }\href {\doibase https://doi.org/10.1016/S0169-5347(99)01664-X}
  {\bibfield  {journal} {\bibinfo  {journal} {Trends in Ecology \& Evolution}\
  }\textbf {\bibinfo {volume} {14}},\ \bibinfo {pages} {361} (\bibinfo {year}
  {1999})}\BibitemShut {NoStop}%
\bibitem [{\citenamefont {Rao}\ \emph {et~al.}(2021)\citenamefont {Rao},
  \citenamefont {Muyinda},\ and\ \citenamefont
  {De~Baets}}]{sodhan2021metapopulation}%
  \BibitemOpen
  \bibfield  {author} {\bibinfo {author} {\bibfnamefont {S.}~\bibnamefont
  {Rao}}, \bibinfo {author} {\bibfnamefont {N.}~\bibnamefont {Muyinda}}, \ and\
  \bibinfo {author} {\bibfnamefont {B.}~\bibnamefont {De~Baets}},\ }\href
  {\doibase https://doi.org/10.1038/s41598-021-93438-8} {\bibfield  {journal}
  {\bibinfo  {journal} {Scientific Reports}\ }\textbf {\bibinfo {volume} {11}}
  (\bibinfo {year} {2021}),\
  https://doi.org/10.1038/s41598-021-93438-8}\BibitemShut {NoStop}%
\bibitem [{\citenamefont {Connolly}\ \emph {et~al.}(2001)\citenamefont
  {Connolly}, \citenamefont {Wayne},\ and\ \citenamefont
  {Bazzaz}}]{connolly2001interspecific}%
  \BibitemOpen
  \bibfield  {author} {\bibinfo {author} {\bibfnamefont {J.}~\bibnamefont
  {Connolly}}, \bibinfo {author} {\bibfnamefont {P.}~\bibnamefont {Wayne}}, \
  and\ \bibinfo {author} {\bibfnamefont {F.~A.}\ \bibnamefont {Bazzaz}},\
  }\href {\doibase https://doi.org/10.1086/318631} {\bibfield  {journal}
  {\bibinfo  {journal} {The American Naturalist}\ }\textbf {\bibinfo {volume}
  {157}},\ \bibinfo {pages} {107} (\bibinfo {year} {2001})}\BibitemShut
  {NoStop}%
\bibitem [{\citenamefont {Majhi}\ \emph {et~al.}(2022)\citenamefont {Majhi},
  \citenamefont {Perc},\ and\ \citenamefont {Ghosh}}]{majhi2022dynamics}%
  \BibitemOpen
  \bibfield  {author} {\bibinfo {author} {\bibfnamefont {S.}~\bibnamefont
  {Majhi}}, \bibinfo {author} {\bibfnamefont {M.}~\bibnamefont {Perc}}, \ and\
  \bibinfo {author} {\bibfnamefont {D.}~\bibnamefont {Ghosh}},\ }\href
  {\doibase doi.org/10.1098/rsif.2022.0043} {\bibfield  {journal} {\bibinfo
  {journal} {Journal of the Royal Society Interface}\ }\textbf {\bibinfo
  {volume} {19}},\ \bibinfo {pages} {20220043} (\bibinfo {year}
  {2022})}\BibitemShut {NoStop}%
\bibitem [{\citenamefont {Buss}\ and\ \citenamefont
  {Jackson}(1979)}]{buss1979competitive}%
  \BibitemOpen
  \bibfield  {author} {\bibinfo {author} {\bibfnamefont {L.}~\bibnamefont
  {Buss}}\ and\ \bibinfo {author} {\bibfnamefont {J.}~\bibnamefont {Jackson}},\
  }\href {\doibase https://doi.org/10.1086/283381} {\bibfield  {journal}
  {\bibinfo  {journal} {The American Naturalist}\ }\textbf {\bibinfo {volume}
  {113}},\ \bibinfo {pages} {223} (\bibinfo {year} {1979})}\BibitemShut
  {NoStop}%
\bibitem [{\citenamefont {Paquin}\ and\ \citenamefont
  {Adams}(1983)}]{paquin1983relative}%
  \BibitemOpen
  \bibfield  {author} {\bibinfo {author} {\bibfnamefont {C.~E.}\ \bibnamefont
  {Paquin}}\ and\ \bibinfo {author} {\bibfnamefont {J.}~\bibnamefont {Adams}},\
  }\href {\doibase https://doi.org/10.1038/306368a0} {\bibfield  {journal}
  {\bibinfo  {journal} {Nature}\ }\textbf {\bibinfo {volume} {306}},\ \bibinfo
  {pages} {368} (\bibinfo {year} {1983})}\BibitemShut {NoStop}%
\bibitem [{\citenamefont {Friedman}\ \emph {et~al.}(2017)\citenamefont
  {Friedman}, \citenamefont {Higgins},\ and\ \citenamefont
  {Gore}}]{friedman2017community}%
  \BibitemOpen
  \bibfield  {author} {\bibinfo {author} {\bibfnamefont {J.}~\bibnamefont
  {Friedman}}, \bibinfo {author} {\bibfnamefont {L.~M.}\ \bibnamefont
  {Higgins}}, \ and\ \bibinfo {author} {\bibfnamefont {J.}~\bibnamefont
  {Gore}},\ }\href {\doibase https://doi.org/10.1038/s41559-017-0109}
  {\bibfield  {journal} {\bibinfo  {journal} {Nature Ecology \& Evolution}\
  }\textbf {\bibinfo {volume} {1}} (\bibinfo {year} {2017}),\
  https://doi.org/10.1038/s41559-017-0109}\BibitemShut {NoStop}%
\bibitem [{\citenamefont {May}(1972)}]{may1972will}%
  \BibitemOpen
  \bibfield  {author} {\bibinfo {author} {\bibfnamefont {R.~M.}\ \bibnamefont
  {May}},\ }\href {\doibase https://doi.org/10.1038/238413a0} {\bibfield
  {journal} {\bibinfo  {journal} {Nature}\ }\textbf {\bibinfo {volume} {238}},\
  \bibinfo {pages} {413} (\bibinfo {year} {1972})}\BibitemShut {NoStop}%
\bibitem [{\citenamefont {Holling}(1973)}]{holling1973resilience}%
  \BibitemOpen
  \bibfield  {author} {\bibinfo {author} {\bibfnamefont {C.~S.}\ \bibnamefont
  {Holling}},\ }\href {\doibase
  https://doi.org/10.1146/annurev.es.04.110173.000245} {\bibfield  {journal}
  {\bibinfo  {journal} {Annual Review of Ecology and Systematics}\ }\textbf
  {\bibinfo {volume} {4}},\ \bibinfo {pages} {1} (\bibinfo {year}
  {1973})}\BibitemShut {NoStop}%
\bibitem [{\citenamefont {Allesina}\ and\ \citenamefont
  {Tang}(2012)}]{allesina2012stability}%
  \BibitemOpen
  \bibfield  {author} {\bibinfo {author} {\bibfnamefont {S.}~\bibnamefont
  {Allesina}}\ and\ \bibinfo {author} {\bibfnamefont {S.}~\bibnamefont
  {Tang}},\ }\href {\doibase https://doi.org/10.1038/nature10832} {\bibfield
  {journal} {\bibinfo  {journal} {Nature}\ }\textbf {\bibinfo {volume} {483}},\
  \bibinfo {pages} {205} (\bibinfo {year} {2012})}\BibitemShut {NoStop}%
\bibitem [{web()}]{web_1}%
  \BibitemOpen
  \href@noop {} {}\bibinfo {howpublished}
  {{\color{blue}https://github.com/CRHENS/HOI-cyclic-interaction}}\BibitemShut
  {NoStop}%
\end{thebibliography}%

%\begin{thebibliography}{46}
%	
%	\bibitem{holme2012temporal} 

%\end{thebibliography}

\end{document}